\documentclass[11pt]{article}  
\usepackage{preamble}
\usepackage{preamble_h_wu}
\usepackage[top=2cm,bottom=2cm,left=2cm,right=2cm,footskip=0.8cm]{geometry}

\title{Joint Communication and Channel Discrimination
\footnotetext{The authors are with the Information and Communication Theory Lab, Department of Electrical Engineering, Eindhoven University of Technology, 5600 MB Eindhoven, The Netherlands (e-mail: h.wu1@tue.nl; h.joudeh@tue.nl).
This work was supported in part by the European Research Council (ERC) through the ERC Starting Grant N. 101116550 (IT-JCAS).}
}
\date{}
\author{Han~Wu and Hamdi~Joudeh}
\begin{document}
\maketitle
\begin{abstract}
We consider a basic joint communication and sensing setup comprising a transmitter, a receiver and a sensor. The transmitter sends a codeword to the receiver through a discrete memoryless channel, and the receiver is interested in decoding the transmitted codeword.
At the same time, the sensor picks up a noisy version of the transmitted codeword through one of two possible discrete memoryless channels.
The sensor knows the codeword and wishes to discriminate between the two possible channels, i.e. to identify the channel that has generated the output given the input.
We study the trade-off between communication and sensing in the asymptotic regime, captured in terms of the channel coding rate against the two types of discrimination error exponents.
We characterize the optimal trade-off between the rate and the exponents for general discrete memoryless channels with an input cost constraint.
\end{abstract}
\section{Introduction}
\label{sec:introduction}
\begin{figure}[h]
\centering
\includegraphics[width=0.7\textwidth]{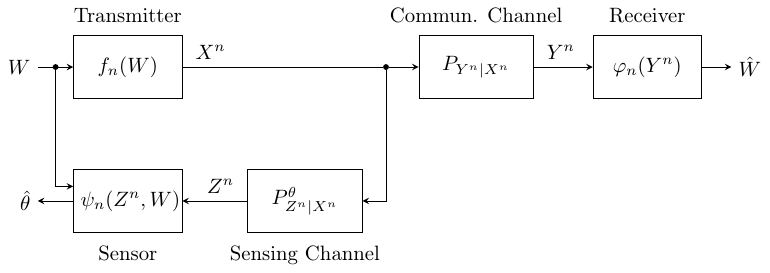}
\caption{An illustration of the considered setting. A precise definition of all blocks is given in Section \ref{sec:problem_setting}.}
\label{fig:block_diagram}
\end{figure}

We consider a setting comprising a transmitter, a receiver and a sensor.
The transmitter has a random message $W$ which it encodes into a sequence $X^n \triangleq X_1,X_2,\ldots,X_n$ of length $n$, drawn from an alphabet $\mathcal{X}^n$. This sequence serves as an input to a pair of channels $P_{Y^n|X^n}: \mathcal{X}^n \to \mathcal{Y}^n$ and $P_{Z^n|X^n}^{\theta} : \mathcal{X}^n \to \mathcal{Z}^n$, where $\mathcal{Y}^n$ and $\mathcal{Z}^n$ are the corresponding output alphabets.
The receiver observes $Y^n \triangleq Y_1,Y_2,\ldots,Y_n$ through $P_{Y^n|X^n}$ and wishes to retrieve the message $W$ from $Y^n$.
The sensor, on the other hand, observes $Z^n \triangleq Z_1,Z_2,\ldots,Z_n$ through $P_{Z^n|X^n}^{\theta}$, which depends on a fixed yet unknown parameter $\theta \in \Theta$.
The sensor has access to $W$ as side information (the transmitter and sensor are, e.g., co-located) and wishes to estimate the channel parameter $\theta$ from $(Z^n,W)$.
An illustration of this setting is shown in Fig. \ref{fig:block_diagram}.

The above setting is a very basic model for joint communication and sensing (JCAS), or integrated sensing and communication (ISAC).
The JCAS/ISAC paradigm has emerged with the aim of designing wireless systems in which transceivers utilize the same hardware and spectrum resources efficiently to both communicate and sense; and has received increased research attention in recent years; see, e.g. \cite{Liu2022,Xiong2024}.
Potential practical use cases include next-generation cellular networks, where base stations will be able to communicate with active devices and simultaneously detect and track passive moving targets from backscattered signals \cite{Wild2021}; and automotive, where vehicles will be able to sense their surroundings to identify and track road obstacles, and communicate with other vehicles on the road and infrastructure \cite{Sturm2011,Ma2020}.

In this paper, our aim is to shed some light on the fundamental performance limits and trade-offs in JCAS systems.
As a step in this direction, we focus on discrete memoryless settings: the input and output alphabets are finite, and the noisy channels are stationary and memoryless.
Moreover, we also limit our attention to the case where the parameter $\theta$ is drawn from $\Theta = \{0,1\}$, which represents the most basic sensing task of target detection or classification.
That is, with knowledge of $W$ (and hence $X^n$) and upon observing $Z^n$, the sensor wishes to discriminate between the two channels $P_{Z^n|X^n}^{0}$ and $P_{Z^n|X^n}^{1}$.

Note that in the setting we consider, the sensing channel parameter (or state) to be estimated does not influence the communication channel. This is an abstraction of practical scenarios where the phenomenon or target to be sensed is distinct from the device or user involved in communication; see, e.g., \cite{Liu2022,Xiong2024}.
\subsection{Related work}
A basic information-theoretic formulation for JCAS was proposed by Kobayashi et al. \cite{Kobayashi2018}, involving a terminal communicating with a second terminal over a state-dependent memoryless channel, and simultaneously estimating the channel state sequence from generalized feedback.\footnote{In relation to the model in Fig. \ref{fig:block_diagram}, the transmit-estimate terminal in \cite{Kobayashi2018} includes both the transmitter and sensor.}
The performance trade-off between communication and state estimation is characterized in terms of a capacity-distortion function, a quantity borrowed from earlier works on state amplification \cite{Sutivong2005,Choudhuri2013,Zhang2011}. 
The results and insights from \cite{Kobayashi2018} were extended by several authors in multiple directions, including multi-terminal settings \cite{Kobayashi2019,Ahmadipour2024,Ahmadipour2023}, secrecy-constrained settings \cite{Ahmadipour2023b,Gunlu2023}, and multi-antenna Gaussian settings \cite{Xiong2023}, to mention a few.
The capacity-distortion trade-off in the original setting of \cite{Kobayashi2018} has also been studied under a special logarithmic loss distortion measure, yielding a simple characterization in terms of mutual information quantities \cite{Joudeh2024}.

All the above-mentioned works follow the same modeling logic in \cite{Kobayashi2018}. That is, the state to be estimated varies in an i.i.d. fashion from one channel use to the other.
This model fails to capture scenarios where the sensing task involves estimating a state (or parameter) that changes at a much slower timescale compared to channel uses.
To study this latter case, one may consider an abstraction where the state to be estimated remains fixed throughout the whole transmission period, a model that sits at the extreme opposite of the i.i.d. state model.
This approach was first taken in by Joudeh and Willems in \cite{Joudeh2022}, where a special case of the setting in Fig. \ref{fig:block_diagram} with discrete binary channels was considered, as well as a case with continuous Gaussian channels.
In both cases studied in \cite{Joudeh2022}, the sensing task considered is that of target detection, i.e. discriminating between a target response and pure noise.
The extension to discriminating between an arbitrary pair of channels in discrete memoryless settings, as described in Fig. \ref{fig:block_diagram}, was considered in our preliminary works \cite{Wu2022,Wu2022arXiv}.
Concurrently, Chang et al. \cite{Chang2022} studied an almost identical model to the one in Fig. \ref{fig:block_diagram}, and further investigated the case of discriminating between more than two channels, and the role of adaptive schemes; a work that was later extended in \cite{Chang2023}.
Other related works include the extension of the target detection setting in \cite{Joudeh2022} to vector Gaussian channels (i.e. multiple antennas) \cite{Joudeh2023}, as well as strong converse results for discrete memoryless settings in \cite{Ahmadipour2023strong,Ahmadipour2024strong}.

Another seemingly related line of work considers the problem of joint detection and decoding at the receiver \cite{Weinberger2014,Weinberger2017}. In the most basic instance of this problem, the receiver wishes to detect the presence of a codeword, i.e. discriminate between codeword and noise, and decode it in case it is present. 
In contrast with the setting in Fig. \ref{fig:block_diagram}, the joint detection and decoding problem co-locates the sensor with the receiver and not the transmitter, and is hence quite distinct from the problem we study in the current paper. 

Most relevant to the present paper are the results of Chang et al. in \cite{Chang2022,Chang2023} and our preliminary results in \cite{Wu2022,Wu2022arXiv}. The nuance differences between these works are further elaborated in light of our contributions in the next subsection.
Before we proceed, we highlight a few more relevant works, some from the classical literature. 
The basic sensing task that we consider, with a binary parameter $\theta$, is a simple binary hypothesis testing problem. 
This is a canonical problem in both statistics and information theory, and notable works that characterize the asymptotic performance limits include those by Chernoff \cite{chernoff1952}, Hoeffding \cite{hoeffding1965}, Shannon-Gallager-Berlekamp \cite{shannonLowerBoundsError1967}, Csisz{\'a}r-Longo \cite{csiszar1971}, and Blahut \cite{blahutHypothesisTestingInformation1974}.
The specific version of the problem that we consider here, where the sensor knows (and may control) the input $X^n$ and wishes to distinguish between two channels $P_{Z^n|X^n}^{0}$ and $P_{Z^n|X^n}^{1}$ from an observation $Z^n$, is also known as channel discrimination \cite{Hayashi2009}, hypothesis testing with feedback \cite{polyanskiy2011_ITA},  or controlled sensing \cite{Nitinawarat2013}.
This problem has been considered in a number of works under various assumptions, including fixed-length non-adaptive transmission in
Shannon-Gallager-Berlekamp \cite{shannonLowerBoundsError1967} and 
Blahut \cite{blahutHypothesisTestingInformation1974}; fixed-length adaptive transmission in  Hayashi \cite{Hayashi2009}, Polyanskiy-Verd{\'u} \cite{polyanskiy2011_ITA} and Nitinawarat et al. \cite{Nitinawarat2013}; and variable-length adaptive transmission in Polyanskiy-Verd{\'u}  \cite{polyanskiy2011_ITA} and Nitinawarat et al. \cite{Nitinawarat2013}. 

In this paper, we consider fixed-length transmission 
focusing mainly on the non-adaptive case (i.e. no feedback).
More importantly, in addition to facilitating channel discrimination at the sensor, the input sequence $X^n$ in our setting must also carry a message to the receiver, which distinguishes our problem from the ones previously considered in the channel discrimination and controlled sensing literature.

\subsection{Contribution and comparison}
\label{subsec:contribution}
We consider the setting illustrated in Fig. \ref{fig:block_diagram} with discrete memoryless channels, a binary parameter $\theta$, and an average input cost constraint; and we study the trade-off between reliable message communication and efficient channel discrimination in the asymptotic regime (i.e., $n \to \infty$).
This trade-off is captured in terms of the channel coding rate against the two channel discrimination error exponents (i.e., the rate-exponent region). Note that the error exponents capture the exponential decay rates of the two types of channel discrimination errors, known as type I and type II errors in the hypothesis testing literature.

In our main result (Theorem \ref{theorem:rate_exponent_region} in Section \ref{section:main_result}), we establish the optimal trade-off between the channel coding rate and the two channel discrimination exponents (i.e., optimal rate-exponent region). 
We also provide insights into the trade-off and demonstrate it through a couple of simple examples.
The proof of our main result is obtained by adapting classical results on binary hypothesis testing, combined with a channel coding argument with constrained input sequences (see Section \ref{sec:channel_discrimination} and Section \ref{sec:comm_disc_trade_off}). 
In the proof we encountered an interesting technical challenge, specifically in showing the converse to the channel coding rate.
The coupling with channel discrimination imposes a constraint on the types of admissible input sequences, i.e. their empirical distributions.
However, unlike additive cost constraints \cite{csiszarInformationTheoryCoding2011,ElGamal2011}, or similarly, constraints imposed by sensing an i.i.d. state sequence subject to an additive distortion measure
\cite{Kobayashi2018,Ahmadipour2024,Zhang2011}; the constraint we encounter here is non-convex in the input sequence type.
This prohibits us from directly applying the common approach of upper bounding the multi-letter mutual information through concavity and Jensen's inequality, which strongly relies on the convexity of the set of input distributions that satisfy the cost (or, similarly, the distortion) constraint.
The approach we take here reduces the problem to upper bounding the rate of a constant-composition code, where all input sequences are of the same type, and uses a slightly more refined analysis of the multi-letter mutual information (see Section \ref{subsec:converse}).\footnote{It is worthwhile noting that our converse proof, which relies on reducing the problem to that of constant-composition codes, first appeared in an earlier version of the current paper \cite{Wu2022arXiv}, posted on arXiv on 15 August 2022. Since then, the exact same approach was also adopted in \cite[Section V.B]{Chang2023}, which first appeared on arXiv on 14 October 2022.}


In Section \ref{section:minimax_and_NP}, we consider two important special cases of our general result in Theorem \ref{theorem:rate_exponent_region}.
In the first case, we consider a minimax error criterion for channel discrimination, where the goal is to minimize the worst of the two types of error, and we characterize the optimal rate-exponent trade-off region in this case (see Theorem \ref{theorem:minimax_region}). 
In the second special case, we adopt a Neyman-Pearson channel discrimination error criterion, where the goal is to minimize one type of error while keeping the other type below a predefined threshold, and we derive the optimal trade-off in this case as well (see Theorem \ref{theorem:NP_region}). This latter case is relevant in many practical applications, e.g. in obstacle detection to avoid road collisions in automotive scenarios, a missed detection is much worse than a false alarm.

As mentioned earlier, a special case of the problem addressed in this paper, with binary channels and an on-off channel parameter (i.e., target detection), was considered in \cite{Joudeh2022} under a minimax channel discrimination error criterion. The results we present here generalize \cite[Theorem 1]{Joudeh2022} to arbitrary discrete memoryless channels with input cost, and to the entire trade-off between the rate and the two types of channel discrimination exponents. 
Other very closely related results were also reported in \cite{Chang2023}, and its preliminary version \cite{Chang2022}.
In these works, the authors consider a setting where $\theta$ is not necessarily binary, but belongs to a finite set; and affects both the communication and sensing channels, and hence the communication problem is of a compound nature. Moreover, they investigate both non-adaptive and adaptive schemes.
Nevertheless, \cite{Chang2022,Chang2023} focus only on the minimax channel discrimination criterion.\footnote{\cite{Chang2022,Chang2023} also do not consider an input cost constraint, which is, however, a minor difference and can be easily incorporated.} 
Our results are somewhat comparable to \cite[Theorem 3]{Chang2022} and \cite[Theorem 1]{Chang2023}, yet are more general in some sense as we characterize trade-off between the two exponents, and are more restricted in another sense as we focus on the binary parameter case.
In this context, it should be noted that for binary $\theta$ and under fixed-length transmission, adaptivity does not improve the channel discrimination error exponents \cite{Hayashi2009,polyanskiy2011_ITA}; and hence there is no loss in generality in our restriction to non-adaptive schemes.

A more subtle difference compared to the preliminary work of Chang et al. \cite{Chang2022} is in the definition of the discrimination error. As we shall see in Section \ref{sec:problem_setting}, we define the two types of discrimination error probabilities by taking the maximum (i.e. worst-case) over all codewords in the codebook. 
We believe this to be a natural definition from an operational perspective, since it provides performance guarantees for sensing regardless of which codeword is used for communication, as it is not known beforehand which codeword (or message) will be selected.
This worst-case formulation, however, also requires more involved analysis. For instance, i.i.d. code ensembles are insufficient for proving achievability in this case, due to the fact that bad codewords from a channel discrimination perspective, while improbable, are still possible under such ensembles.
In \cite{Chang2022}, the authors alleviate this challenge by considering the average discrimination error over all codewords, which provides no guarantees on the sensing performance for the worst-case codeword, but renders i.i.d. code ensembles sufficient. 
In our preliminary work \cite{Wu2022arXiv}, we dealt with worst-case sensing errors by resorting to constant composition codes (see also \cite{Joudeh2022}), which were also later adopted by Chang et al. in \cite{Chang2023}.
In the current paper, we prove achievability using almost constant-composition codes through a constrained version of the channel coding theorem, which has the advantage of being directly applicable to channels with continuous alphabets.
\subsection{Notation}
Upper-case letters, e.g. $X,Y,Z$, often denote random variables and the corresponding lower-case letters, e.g. $x,y,z$, denote their realizations. Calligraphic letters, e.g. $\mathcal{M}$, denote sets. $|\mathcal{M}|$ denotes the cardinality of  set $\mathcal{M}$. The indicator function $ \ind\left[\mathcal{A} \right]$ is equal to $1$ if the event $\mathcal{A}$ is true, and $0$ otherwise. Let $X$ and $Y$ be respectively  an input and output  to a channel $P_{Y|X}$, which is a stochastic mapping from the input alphabet $\mathcal{X}$ to the output alphabet $\mathcal{Y}$.
The mutual information $I(X;Y)$ is also denoted by $I(P_X, P_{Y|X})$.
The Bernoulli distribution with parameter $p$ is denoted by $\Bern(p)$ and the binary symmetric channel with parameter $q$ is denoted by $\bsc(q)$.
For  $p,q \in [0,1]$, we define $ p \ast q \triangleq (1-q)p + q(1-p) $.

Next, we present some notation and preliminaries on types from \cite{csiszarInformationTheoryCoding2011}, which will be essential in the technical development of our results.
Given a sequence  \(x^{n} \in \mathcal{X}^{n}\), we define
\begin{equation}
    N(a|x^{n}) \triangleq \sum_{i=1}^{n} \ind \left[ x_i = a \right], a \in \mathcal{X}.
\end{equation}
The type of \(x^{n}\), denoted by \(\hat{P}_{x^{n}}\), is  a distribution on \(\mathcal{X}\) defined as
\begin{equation}
    \hat{P}_{x^{n}} (a)= \frac{N(a|x^{n})}{n}, \ a \in \mathcal{X}.
\end{equation}
Let \(\mathcal{P}(\mathcal{X})\) be the set of all distributions (i.e. probability mass functions) on $\mathcal{X}$ and let \(\mathcal{P}_n(\mathcal{X})\) be the set of all types of sequences in $\mathcal{X}^n$. Note that \(\mathcal{P}_n(\mathcal{X}) \subset \mathcal{P}(\mathcal{X})\). A very important property is that the number of types in \(\mathcal{P}_n(\mathcal{X}) \) is at most a polynomial in $n$, which follows from the upper bound \cite[Lemma 2.2]{csiszarInformationTheoryCoding2011}
\begin{equation}
\label{eq:type_counting_bound}
\abs{\mathcal{P}_n(\mathcal{X})}  \leq (n+1)^{\abs{\mathcal{X}}}.
\end{equation}
\section{Problem Setting}
\label{sec:problem_setting}
We consider the setting introduced in Section \ref{sec:introduction} and illustrated in Fig. \ref{fig:block_diagram} with finite alphabets $\mathcal{X}$, $\mathcal{Y}$, $\mathcal{Z}$ and a binary parameter $\theta \in \{0,1 \}$.
The channels are  stationary and memoryless, that is
\begin{equation}
\label{eq:DMCs}
P_{Y^n|X^n}( y^n | x^n)  = \prod_{i = 1}^{n}  P_{Y|X}( y_i | x_i) \quad \text{and}
\quad   P_{Z^n|X^n}^{\theta}( z^n | x^n)  = \prod_{i = 1}^{n}  P_{Z|X}^{\theta}( z_i | x_i)
\end{equation}
where $n$ is a positive integer that denotes the block length (i.e. number of channel uses).
An admissible input sequences $x^n \in \mathcal{X}^n$ must satisfy an average cost constraint of
\begin{equation}
\label{eq:cost_constraint}
    \frac{1}{n} \sum_{i = 1}^{n} b(x_i) \leq B
\end{equation}
where  $b:\mathcal{X} \to \mathbb{R}_{+}$ is some non-negative cost function and $B \geq 0$ is the average cost constraint.

To simplify the notation, we use $P_0^n(z^{n}|x^{n}) = \prod_{i=1}^{n}P_0(z_i | x_i)$  and $P_1^n(z^{n}|x^{n}) = \prod_{i=1}^{n}P_1(z_i | x_i)$  to denote $P_{Z^{n} | X^{n}}^{0}(z^{n} | x^{n})$ and  $P_{Z^{n} | X^{n}}^{1}(z^{n} | x^{n})$ respectively,  where $P_0$ and $P_1$ denote $P_{Z|X}^0$ and $P_{Z|X}^1$ respectively. Throughout the paper, we assume that \(P_0(z|x)P_1(z|x) \neq 0\), for all $x \in \mathcal{X}$ and $z \in \mathcal{Z}$, i.e., the two channels have a shared support under every input. This mild regularity condition helps avoid unnecessary technical complications, and is satisfied from many channels of interest.
\subsection{Codes and error probabilities}
\label{subsec:codes}
For fixed block length $n$, let $\mathcal{M}_n \triangleq \big\{1,2,\ldots, M_n \big\}$ be a message set of $M_n$ message indices. An $(n,M_n)$-code for the above setting
consists of the following mappings:
\begin{itemize}
\item An encoding function $f_n : \mathcal{M}_n \to \mathcal{X}^n$ that maps each message index $w \in \mathcal{M}_n$ into a  codeword $x^n(w) = f_n(w)$ from $\mathcal{X}^n$, that satisfies the cost constraint in \eqref{eq:cost_constraint}. The corresponding set of all $M_n$ codewords, given by $\mathcal{C}_n \triangleq \{x^n(1),x^n(2),\ldots,x^n(M_n)\}$, is known as a codebook.
\item A message decoding function  $\varphi_n : \mathcal{Y}^n \to \mathcal{M}_n$ that maps each output sequence $y^n \in \mathcal{Y}^n $  into  a decoded message $\hat{w} = \varphi_n(y^n)$ from the message set $\mathcal{M}_n$.
\item A channel discrimination function $\psi_n : \mathcal{Z}^n \times \mathcal{M}_n \to \{0,1\} $ that maps each output sequence and message pair $(z^n,w) \in \mathcal{Z}^n \times \mathcal{M}_n$ into a decision (i.e. a hypothesis)  $\hat{\theta} = \psi_n(z^n,w)  $ from $\{0,1\}$.
\end{itemize}

The message $W$, which is drawn uniformly at random from $\mathcal{M}_n $, is encoded into $X^n = f_n(W)$ and then sent over the channels.
Upon observing $Y^n$, the receiver produces a decoded message $\hat{W} = \varphi_n(Y^n)$.
On the other end, upon observing $Z^n$ and with knowledge of $W$, the sensor produces a binary decision $\hat{\theta} = \psi_n(Z^n,W)$.
Next, we examine the message decoding and channel discrimination error probabilities.

\emph{Decoding error:} For a given code, the probability of decoding error given that message $W = w$ has been sent is $\Prb\left[ \varphi_n(Y^n) \neq w \mid X^n = x^n(w) \right]$.
The maximum probability of decoding error is defined as
\begin{equation}
    p_{\mathrm{e},n} \triangleq \max_{w \in \mathcal{M}_n} \Prb\left[ \varphi_n(Y^n) \neq w \mid X^n = x^n(w) \right] \label{eq:max_decoding_error}
\end{equation}
which is a common performance measure that reflects the assumption that messages are equally important.
It is sometimes more convenient to work with the average probability of decoding error, define as
\begin{equation}
    p_{\mathrm{e},n}^{\mathrm{av}} \triangleq \frac{1}{M_n}\sum_{w \in \mathcal{M}_n} \Prb\left[ \varphi_n(Y^n) \neq w \mid X^n = x^n(w) \right]. \label{eq:av_decoding_error}
\end{equation}
As is often the case with asymptotic rate results in DMCs, the results we present here remain unchanged regardless of whether we choose  $ p_{\mathrm{e},n} $ or $ p_{\mathrm{e},n}^{\mathrm{av}}$ as the measure of decoding error probability.
In some cases, where it helps to emphasize the codebook being used, we write
$p_{\mathrm{e},n}(\mathcal{C}_n)$  and $p_{\mathrm{e},n}^{\mathrm{av}}(\mathcal{C}_n)$.

\emph{Discrimination error:}
Without loss of generality, any discrimination function can be written as 
\begin{equation}
    \psi_n(z^n,w) = \ind \left[ z^n \notin \mathcal{A}_n(w) \right]
\end{equation}
for some decision region\footnote{Without loss of generality, we prohibit the trivial degenerate cases of $\mathcal{A}_n(w) = \emptyset$ and $\mathcal{A}_n(w) = \mathcal{Z}^n$.} $\mathcal{A}_n(w) \subset \mathcal{Z}^n$, comprising output sequences that map to hypothesis $\hat{\theta} = 0$.
There are two types of discrimination error events, associated with the two hypotheses, defined for $W = w$ as 
\begin{align}
\label{eq:discrimination_error_probabilities}
\varepsilon_{0,n}(w) & \triangleq   \Prb \left[ Z^n \notin \mathcal{A}_n(w)  \mid  \theta = 0, X^n = x^n(w)  \right] \\
\varepsilon_{1,n}(w)  & \triangleq  \Prb \left[ Z^n \in \mathcal{A}_n(w) \mid   \theta = 1, X^n = x^n(w)  \right]
\end{align}
known as the type I error and type II error, respectively. 
Since it is not known beforehand which message will be sent, it is reasonable to define the discrimination error probabilities for a code  as
\begin{align}
\label{eq:two_error_definition}
    \varepsilon_{0,n} & = \max_{w \in \mathcal{M}_n} \varepsilon_{0,n}(w)\\
    \varepsilon_{1,n} & = \max_{w \in \mathcal{M}_n} \varepsilon_{1,n}(w).
\end{align}
This worst-case definition guarantees that by considering the pair $(\varepsilon_{0,n}, \varepsilon_{1, n})$, a certain discrimination error performance is guaranteed regardless of which messages has been selected.  

\subsection{Rate-Exponent region}
We are interested in the asymptotic performance limits measured in terms of the channel coding rate and the channel discrimination error exponents. This is formalized as follows.
\begin{definition}
\label{definition:R_E_region}
A rate-exponent tuple $(R,E_0,E_1)$ is said to be achievable if for every $\epsilon > 0$, and provided that $n \geq n_{\epsilon}$ for some (possibly large) $n_{\epsilon}$, there exists an $(n,M_n)$-code with
\begin{align}
p_{\mathrm{e},n} & \leq \epsilon \\
\frac{1}{n} \log  M_n  & \geq  R - \epsilon\\
-\frac{1}{n}\log \varepsilon_{0,n} & \geq E_0 - \epsilon 
\label{eq:discr_error_exp_0}\\
-\frac{1}{n}\log  \varepsilon_{1,n}  & \geq  E_1 - \epsilon.
\label{eq:discr_error_exp_1}
\end{align}
The rate-exponent region $\mathcal{R}$ is the closure of the set of all achievable tuples $(R,E_0, E_1)$.
\end{definition}
The main result of this paper is a characterization of the rate-exponent region $\mathcal{R}$ for the general discrete memoryless channels in \eqref{eq:DMCs} under the input cost constraint in \eqref{eq:cost_constraint}.
\section{Main Result}
\label{section:main_result}
We now present the main result of the paper, which is a complete characterization of the rate-exponent region in Definition \ref{definition:R_E_region}.
To this end, we define the channel \(P_s: \mathcal{X} \to \mathcal{Z}\) for $s\in [0,1]$ as
\begin{equation}
    P_s (z|x) = \frac{P_0(z|x)^{1-s} P_1(z|x)^{s}}{\sum_{z' \in \mathcal{Z}}P_0(z'|x)^{1-s} P_1(z'|x)^{s}}.
\end{equation}
For any fixed input $x$, \(P_s(\cdot|x)\) is a parameterized output distribution on $\mathcal{Z}$ known in the literature as the ``tilted'' distribution, which moves from \(P_0(\cdot|x)\) to \(P_1(\cdot|x)\) as the parameter \(s\) increases.
We also have the following standard definition of the conditional relative entropy (or conditional information divergence) between any pair of channels $P_{Z|X}$ and $Q_{Z|X}$ from $\mathcal{X}$ to $\mathcal{Z}$ given an input distribution $P_{X}$ on $\mathcal{X}$:
\begin{equation}
\nonumber
    D \left( P_{Z|X} \| Q_{Z|X} | P_X \right) = \sum_{x \in \mathcal{X}} P_X(x)  D \left( P_{Z|X}(\cdot|x) \| Q_{Z|X} (\cdot|x) \right) = \sum_{x \in \mathcal{X}} P_X(x) \sum_{z \in \mathcal{Z}} P_{Z|X}(z|x) \log \frac{P_{Z|X}(z|x)}{Q_{Z|X}(z|x)}.
\end{equation}
We assume that $P_{Z|X}(z|x) = 0$ whenever $Q_{Z|X}(z|x) = 0$, and hence  
$D(P_{Z|X} \| Q_{Z|X} | P_X) < \infty$. This is satisfied for all channels of interest to us here. 
We are now ready to present the main result. 
\begin{theorem}
\label{theorem:rate_exponent_region}
\(\mathcal{R}\) is given by the set of all non-negative pairs \((R,E_0, E_1)\) such that
\begin{align}
    R &\leq I(P_X, P_{Y|X}) \\
    E_0 &\leq D(P_s\|P_0 | P_X) \\
    E_1 &\leq D(P_s\|P_1 | P_X)
\end{align}
for some input distribution \(P_X \in \mathcal{P}(\mathcal{X})\) that satisfies \(\E_{X \sim P_X}[b(X)] \leq B\), and some \(s \in [0,1]\).
\end{theorem}
The proof of the above theorem is presented in the Section \ref{sec:comm_disc_trade_off}. As a first step towards the proof,  we first study the trade-off between the two discrimination exponents $E_0$ and $E_1$ in Section \ref{sec:channel_discrimination}.
Moreover, two important special cases of the result in Theorem \ref{theorem:rate_exponent_region}, where we specialize to minimax and Neyman-Pearson discrimination error metrics, are presented and discussed in Section \ref{section:minimax_and_NP}.
\subsection{Insights and examples}
We now provide some insights into the result in Theorem \ref{theorem:rate_exponent_region}. To this end, it helps to interpret the input distribution $P_X$ in Theorem \ref{theorem:rate_exponent_region} as being equal, or arbitrarily close, to the type (i.e. empirical distribution) of codewords in the employed codebooks.
Under such codebooks, classical results on binary hypothesis testing are used to establish that the discrimination errors are given by
\begin{equation}
  \varepsilon_{0,n} \approx e^{-n D(P_s\|P_0 | P_X)} \text{ and }  \varepsilon_{1,n} \approx e^{-n D(P_s\| P_1 | P_X)}.
\end{equation}
for some parameter $s \in [0,1]$, which is used to tune the trade-off between the two types of error probabilities, and is directly related to the choice of channel discrimination function (or decision region).  
On the other hand, a channel coding argument with constrained input sequences is used to show that reliable communication is achieved provided that the number of codewords in the codebook is $M_n \approx e^{n I(P_X, P_{Y|X})} $.

As seen through the above explanation, and as one may expect, the type $P_X$ of codewords dictates the performance of both communication and channel discrimination. 
In general, a trade-off between the two arises since $P_X$ that maximizes the rate may not simultaneously provide the best discrimination exponents.
This trade-off is further explored through the following basic examples. 
\begin{example}
\label{example:1}
Consider a setting with a binary input, and binary outputs given by
\begin{equation}
Y  = X \oplus N_{Y}  \ \text{and} \ Z  = \theta X \oplus N_{Z}  
\end{equation}
where $N_{Y}$ and $N_{Z}$ are Bernoulli with parameters $p$ and $q$ respectively, and $\theta \in \{0,1\}$.
Here $P_{Y|X}$ is a $\bsc(p)$, $P_0$ is a $\bsc(q)$, while $P_1$ satisfies $P_1(1|\cdot) = q$ and $P_1(0|\cdot)=1-q$.
Therefore, we have
\begin{align}
    & P_s (0|0) = 1 - P_s (1|0) = 1-q \\
    & P_s (0|1) = 1 - P_s (1|1) = \frac{(1-q)^{1-s}q^s}{(1-q)^{1-s}q^s + q^{1-s}(1-q)^s} = \hat{q}
\end{align}
where  $\hat{q} \in [\min\{q,1-q\} , \max\{q,1-q\}]$, depending on $s$.
Let $P_X \sim \Bern (\rho)$. It follows that
\begin{align}
    I(P_X, P_{Y|X}) &= H(\rho * p) -H(p) \\
    D(P_s \| P_0 | P_X) &= \rho d(\hat{q} \| 1-q) \\
    D(P_s \| P_1 | P_X) &= \rho d(\hat{q} \| q)
\end{align}
where $H(p)$ is the entropy of the distribution $\Bern(p)$, while $d(q\| 1-q)$ is the information divergence between $\Bern(q)$ and $\Bern(1-q)$.
From the above, it follows $\mathcal{R}$ for this setting is described by
\begin{align}
     R   &\leq H(\rho * p) -H(p) \\
     E_0 &\leq \rho d(\hat{q} \| 1-q) \\
     E_1 &\leq \rho d(\hat{q} \| q)
\end{align}
for some $\hat{q} \in [\min\{q,1-q\} , \max\{q,1-q\}]$ and $\rho \leq B$ (we assume  $b(1) = 1$ and $B \leq 1$).
The maximum rate is achieved when $\rho= \min\{0.5,B\}$, while the best set of exponent pairs is achieved when $\rho = B$. Hence, there is a trade-off between the rate on one hand and the pair of exponents on the other whenever $B > 0.5$. Note that $\hat{q}$ controls the exponents trade-off for any fixed $\rho$, and does not influence the rate. 
\end{example}
\begin{example}
\label{example:2}
Consider a binary input binary output settings as in the previous example, but here  $P_0$ is a $\bsc(p_0)$ and $P_1$ is a $\bsc(p_1)$. In this case, for any $s \in [0,1]$ we have
\begin{equation}
    P_s (0|0) =  P_s (1|1) = \frac{(1-p_0)^{1-s} (1-p_1)^{s}}{(1-p_0)^{1-s} (1-p_1)^{s} + p_0^{1-s} p_1^{s}} = 1 - q
\end{equation}
where $ q  \in [\min\{p_0,p_1\}, \max\{p_0,p_1\} ]$. It follows that 
\begin{align}
    D(P_s \| P_0 | P_X) &= d(q \| p_0) \\
    D(P_s \| P_1 | P_X) &= d(q \| p_1)
\end{align}
which do not depend on the input cost constraint. Therefore, $\mathcal{R}$ in this case is given by 
\begin{align}
    R &\leq  H\left( \min\{0.5,B\} * p \right) - H(p) \\
    E_0 &\leq d(q\| p_0) \\
    E_1 &\leq d(q \| p_1)
\end{align}
for some  $ q  \in [\min\{p_0,p_1\}, \max\{p_0,p_1\} ]$.
Hence in this case, the rate $R$ and the exponent pair $(E_0,E_1)$ are not coupled, and there is no trade-off between the two tasks.
\end{example}
\section{Channel Discrimination}
\label{sec:channel_discrimination}
In this section, we focus on the problem of channel discrimination with the aid of a fixed input sequence $x^n \in \mathcal{X}^n$, also known as controlled sensing \cite{Nitinawarat2013}. We review classical results from the literature, mainly due to Shannon, Gallager and Berlekamp \cite{shannonLowerBoundsError1967}, and adapt them to the setting considered here. 

With knowledge of $x^n$, channel discrimination boils down to simple binary hypothesis testing between $P^n_0( \cdot | x^n)$ and $P^n_1( \cdot | x^n)$.
It is known that the likelihood ratio test (LRT) is the optimal (deterministic) test in this setting. The decision region can be written in terms of the log-likelihood ration (LLR) as\footnote{We omit the dependency of the decision region and error probabilities on the message index $w$ in this part, as we focus on only one transmitted sequence. This is equivalent to having a codebook of size $M_n = 1$.} 
\begin{equation}
\label{eq:LLR_threshold}
\mathcal{A}_{n}= \left\{ z^n \in \mathcal{Z}^{n} : \log \frac{P_1^n(z^{n}|x^n)}{P_0^n(z^{n}|x^n)} \leq \gamma_n \right\}
\end{equation}
for some threshold $\gamma_n \in \mathbb{R}$, that can be tuned to achieve different trade-offs between the two types of error probability. 
Now suppose that the channel is used once, i.e. $n=1$. In analyzing the error probabilities, the following function, defined on $s \in [0,1]$ for every input symbol $x \in \mathcal{X}$, is instrumental 
\begin{equation}
\label{eq:mu_x}
    \mu_x(s) = \log \sum_{z \in \mathcal{Z}}P_0(z|x)^{1-s} P_1(z|x)^{s}.
\end{equation}
$\mu_x(s) $ is the cumulant-generating function of the LLR $\log \frac{P_1(Z|x)}{P_0(Z|x)}$, evaluated under $Z \sim P_0(\cdot | x)$.
The first and second derivatives of $\mu_x(s)$, denoted by $\mu_x'(s)$ and $\mu_x''(s)$ respectively, are given by
\begin{align}
\label{eq:mu_x_prime}
    \mu^{\prime}_x(s) & =  \sum_{z \in \mathcal{Z}}P_s(z|x) \log \frac{P_1(z|x)}{P_0(z|x)} \\ 
\label{eq:mu_x_prime_prime}    
    \mu^{\prime \prime}_x(s) & =  \sum_{z \in \mathcal{Z}}P_s(z|x) \left(\log \frac{P_1(z|x)}{P_0(z|x)}\right)^2 - \left( \mu^{\prime}_x(s) \right)^2
\end{align}
which coincide, respectively,  with the mean and variance of \( \log \frac{P_1(Z|x)}{P_0(Z|x)}\), evaluated under \(Z \sim P_s(\cdot | x)\).
It follows that \(\mu^{\prime \prime}_x(s) \geq 0\) always holds.
Moreover, it can be verified that $\mu^{\prime \prime}_x(s) > 0$ unless $\frac{P_1(z|x)}{P_0(z|x)}$ is a constant over all $z$. Since we have assumed that the two channels have the same support, this would occur only if they are identical.
Otherwise, if $P_1(\cdot|x)$ and $P_0(\cdot|x)$ are distinct, then $\mu_x(s)$ is a strictly convex function on $s \in [0,1]$.
The following identities can be verified from the expressions in \eqref{eq:mu_x} and \eqref{eq:mu_x_prime}
\begin{align}
\label{eq:div_identity_x_1}
    D\left(P_s(\cdot|x) \| P_0(\cdot|x) \right) &= s\mu_x^{\prime}(s) - \mu_x(s) \\
\label{eq:div_identity_x_2}    
     D\left(P_s(\cdot|x) \| P_1(\cdot|x) \right) &= (s-1)\mu_x^{\prime}(s)-\mu_x(s).
\end{align}

When the channel is used multiple times, in which a sequence $x^n$ is sent over $n$ channel uses, multi-letter extensions of the above quantities become relevant. We denote these by $\mu_{x^n}(s)$, $\mu_{x^n}^{\prime}(s)$ and $\mu_{x^n}^{\prime \prime}(s)$.
Since the channels $P^n_0( \cdot | x^n)$ and $P^n_1( \cdot | x^n)$ are memoryless, it can be verified that 
\begin{align}
\mu_{x^n}(s) & = \sum_{i=1}^n \mu_{x_i}(s) = n \sum_{x \in \mathcal{X}} \hat{P}_{x^n}(x) \mu_{x}(s) \\
\mu_{x^n}^{\prime}(s) & = \sum_{i=1}^n \mu^{\prime}_{x_i}(s) = n \sum_{x \in \mathcal{X}} \hat{P}_{x^n}(x) \mu^{\prime}_{x}(s) 
\\
\mu_{x^n}^{\prime \prime}(s) & = \sum_{i=1}^n \mu^{\prime \prime}_{x_i}(s) = n \sum_{x \in \mathcal{X}} \hat{P}_{x^n}(x) \mu^{\prime \prime}_{x}(s).
\end{align}
The first two identities follow directly from the corresponding definitions, while noting that the tilted channel $P^n_0( \cdot | x^n)$ is memoryless for every $s$.
The last identity is obtained by noting that $\mu_{x^n}^{\prime \prime}(s)$ is the variance of a sum of independent random variables and, therefore, is equal to the sum of the individual variances.
Combining the first two identities with \eqref{eq:div_identity_x_1} and \eqref{eq:div_identity_x_2}, it follows that 
\begin{align}
\label{eq:div_identity_Px_1}
    nD(P_s \|P_0| \hat{P}_{x^n} ) &= s\mu_{x^n}^{\prime}(s) - \mu_{x^n}(s) \\
\label{eq:div_identity_Px_2} 
    nD(P_s \| P_1|\hat{P}_{x^n} ) &= (s-1)\mu_{x^n}^{\prime}(s)-\mu_{x^n}(s).
\end{align}
These conditional divergence terms are useful for writing bounds on the two types of error probabilities.
\begin{lemma}
\label{lemma:HT_error_bounds}
Fix $n$ and $x^n \in \mathcal{X}^n$. There exists a decision region $\mathcal{A}_n \subset \mathcal{Z}^n$ and $s \in (0,1)$ such that
    \begin{align}
    \label{eq:error_UB_0}
      \varepsilon_{0,n} & \leq \exp\left\{ -n D(P_s \| P_0 | \hat{P}_{x^n})\right\} \\
    \label{eq:error_UB_1}
      \varepsilon_{1,n} & \leq \exp\left\{ -n D(P_s \| P_1 | \hat{P}_{x^n})\right\}
    \end{align}
Moreover, for every decision region $\mathcal{A}_n \subset \mathcal{Z}^n$ and $s \in (0,1)$, at least one of the inequalities
    \begin{align}
     \label{eq:error_LB_0}
      \varepsilon_{0,n}& > \frac{1}{4} \exp\left\{ -n D(P_s \| P_0 | \hat{P}_{x^n}) - s\sqrt{2 \mu^{\prime \prime}_{x^n}(s)} \right\} \\
    \label{eq:error_LB_1}
      \varepsilon_{1,n} & > \frac{1}{4} \exp\left\{ -n D(P_s \| P_1 | \hat{P}_{x^n})  - (1-s)\sqrt{2   \mu^{\prime \prime}_{x^n}(s)} \right\}
    \end{align}
must hold.   
\end{lemma}
The above lemma essentially follows from \cite[Theorem 5]{shannonLowerBoundsError1967} by Shannon, Gallager and Berlekamp. We find the proof quite insightful, and hence we present a version that is adapted to our setting next. 
\begin{proof}
Fix a parameter $s \in (0,1)$. We choose the threshold in  \eqref{eq:LLR_threshold} such that 
\begin{equation}
    \mathcal{A}_{n}= \left\{ z^n \in \mathcal{Z}^{n} : \log \frac{P_1^n(z^{n}|x^n)}{P_0^n(z^{n}|x^n)} \leq \mu_{x^n}^{\prime}(s) \right\}.
\end{equation}
It follows that the first error probability can be bounded above as 
\begin{align}
    \varepsilon_{0,n} &=  \sum_{z^{n} \in \mathcal{A}_n^{c}} P_0^n(z^{n} | x^{n})\\
    &=\sum_{z^{n} \in \mathcal{A}_n^{c}}\exp \left\{ \mu_{x^n}(s) - s \log \frac{P_1^n(z^{n}|x^n)}{P_0^n(z^{n}|x^n)} \right\}P_s^n (z^{n}|x^{n}) \\
    &\leq \sum_{z^{n} \in \mathcal{A}_n^{c}}\exp \left\{ \mu_{x^n}(s) - s\mu_{x^n}^{\prime}(s)  \right\}P_s^n (z^{n}|x^{n}) \\
    &\leq \exp \left\{ \mu_{x^n}(s) - s\mu_{x^n}^{\prime}(s)  \right\}.
\end{align}
The main step in the obtaining the above bound is the change of measure trick in the second line. 
Similarly, we bound the second error probability as 
\begin{align}
    \varepsilon_{1,n} &=  \sum_{z^{n} \in \mathcal{A}_n} P_1^n(z^{n} | x^{n})\\
    &=\sum_{z^{n} \in \mathcal{A}_n}\exp \left\{ \mu_{x^n}(s) + (1-s) \log \frac{P_1^n(z^{n}|x^n)}{P_0^n(z^{n}|x^n)} \right\}P^n_s (z^{n}|x^{n}) \\
    &\leq \sum_{z^{n} \in \mathcal{A}_n}\exp \left\{ \mu(s) + (1-s)\mu_{x^n}^{\prime}(s)  \right\}P^n_s (z^{n}|x^{n}) \\
    &\leq \exp \left\{ \mu_{x^n}(s) + (1-s) \mu_{x^n}^{\prime}(s)  \right\}.
\end{align}
The upper bounds in Lemma \ref{lemma:HT_error_bounds} follow by substituting \eqref{eq:div_identity_Px_1} and \eqref{eq:div_identity_Px_2} into the above bounds. 

We now turn to proving the lower bounds. To this end, fix a parameter $s \in (0,1)$ and define the following subset of output sequences
\begin{equation}
  \mathcal{D}_n(s)  = \left\{ z^n \in \mathcal{Z}^{n} : \left| \log \frac{P_1^n(z^{n}|x^n)}{P_0^n(z^{n}|x^n)} - 
      \mu_{x^n}^{\prime}(s) \right| \leq \sqrt{2 \mu_{x^n}^{\prime \prime}(s)} \right\}.
\end{equation}
For an arbitrary decision region $\mathcal{A}_n \subset \mathcal{Z}^n$, the error probability of the first type is bounded below as 
\begin{align}
    \varepsilon_{0,n} & \geq \sum_{z^{n} \in \mathcal{A}_n^{c} \cap \mathcal{D}_n(s) } P_0^n(z^{n} | x^{n})\\
    &=\sum_{z^{n} \in \mathcal{A}_n^{c} \cap \mathcal{D}_n(s) }\exp \left\{ \mu_{x^n}(s) - s \log \frac{P_1^n(z^{n}|x^n)}{P_0^n(z^{n}|x^n)} \right\}P_s^n (z^{n}|x^{n}) \\
    &\geq \exp \left\{ \mu_{x^n}(s) - s\mu_{x^n}^{\prime}(s)  
 - s\sqrt{2 \mu_{x^n}^{\prime \prime}(s)} \right\} \sum_{z^{n} \in \mathcal{A}_n^{c} \cap \mathcal{D}_n(s) } P_s^n (z^{n}|x^{n})
\end{align}
where in the second line, we use the change of measure argument used in the upper bound, while the last inequality follows from the definition of $\mathcal{D}_n(s)$.
Similarly, we bound the second error probability as
\begin{align}
    \varepsilon_{1,n} &=  \sum_{z^{n} \in \mathcal{A}_n \cap \mathcal{D}_n(s) } P_1^n(z^{n} | x^{n})\\
    &=\sum_{z^{n} \in \mathcal{A}_n \cap \mathcal{D}_n(s) }\exp \left\{ \mu_{x^n}(s) + (1-s) \log \frac{P_1^n(z^{n}|x^n)}{P_0^n(z^{n}|x^n)} \right\}P_s^n (z^{n}|x^{n}) \\
    &\geq \exp \left\{ \mu_{x^n}(s) + (1-s)\mu_{x^n}^{\prime}(s)  
 - (1-s)\sqrt{2 \mu_{x^n}^{\prime \prime}(s)} \right\} \sum_{z^{n} \in \mathcal{A}_n \cap \mathcal{D}_n(s) } P_s^n (z^{n}|x^{n}).
\end{align}
We now proceed by observing that the probability of $\mathcal{D}_n(s)$ under $P_s^n(\cdot|x^n)$ is lower bounded as follows
\begin{align}
P_s^n \left( \mathcal{D}_n(s) |x^n \right) = 1 - \Prb_{Z^n \sim P_s^n(\cdot|x^n)} \left[ \left| \log \frac{P_1^n(Z^{n}|x^n)}{P_0^n(Z^{n}|x^n)} - 
      \mu_{x^n}^{\prime}(s) \right|^2 > 2 \mu_{x^n}^{\prime \prime}(s) \right] > \frac{1}{2}
\end{align}
where we have used Markov's (or Chebyshev's) inequality, combined with the fact that $\mu_{x^n}^{\prime}(s)$ and $\mu_{x^n}^{\prime \prime}(s)$ are the mean and variance of $\log \frac{P_1^n(Z^{n}|x^n)}{P_0^n(Z^{n}|x^n)} $ under $Z^n \sim P_s^n(\cdot|x^n)$.
Therefore, it follows that
\begin{equation}
  P_s^n \left( \mathcal{D}_n(s) \cap  \mathcal{A}_n^{c} |x^n \right)   +   P_s^n \left( \mathcal{D}_n(s) \cap  \mathcal{A}_n |x^n \right)  > \frac{1}{2}.
\end{equation}
which in turn implies that either  $P_s^n \left( \mathcal{D}_n(s) \cap  \mathcal{A}_n^{c} |x^n \right)  > \frac{1}{4}$ or
$P_s^n \left( \mathcal{D}_n(s) \cap  \mathcal{A}_n |x^n \right)  > \frac{1}{4}$ must hold.
Combining this with the above lower bounds on the error probabilities completes the proof.
\end{proof}
We conclude this section by rewriting the result in Lemma \ref{lemma:HT_error_bounds} in a form that will be more useful for us when proving Theorem \ref{theorem:rate_exponent_region}.
To this end, and for fixed $n$ and $x^n \in \mathcal{X}^n$, define the region 
\begin{equation}
    \mathcal{E}_n(x^n) = \left\{ (E_{0,n},E_{1,n}) : E_{0,n} = -\frac{1}{n}\log \varepsilon_{0,n} \ \text{and} \ E_{1,n} = - \frac{1}{n}\log \varepsilon_{1,n}, \ \text{for some} \ \mathcal{A}_n \subset \mathcal{Z}^n    \right\}.
\end{equation}
This can be thought of as a non-asymptotic error exponent trade-off region for a fixed input sequence $x^n$. 
An inner bound and outer bound for this region are obtained from Lemma \ref{lemma:HT_error_bounds} as follows. 
\begin{corollary}
\label{corollary:E_trade_off}
$\mathcal{E}_n(x^n)$ includes the region given by all non-negative tuples  $(E_{0,n},E_{1,n})$ satisfying
    \begin{align}
    \label{eq:E_inner_bound_0}
        E_{0,n} &\leq   D(P_s\| P_0 | \hat{P}_{x^n}) \\
    \label{eq:E_inner_bound_1}   
        E_{1,n} &\leq   D(P_s\| P_1 | \hat{P}_{x^n})
    \end{align}
    for some $s \in (0,1)$, and is included in the the region given by all non-negative tuples satisfying
    \begin{align}
    \label{eq:E_outer_bound_0}
    E_{0,n} &\leq   D(P_s\| P_0 | \hat{P}_{x^n}) + \frac{c}{\sqrt{n}} \\
    \label{eq:E_outer_bound_1}
    E_{1,n} &\leq   D(P_s\| P_1 | \hat{P}_{x^n}) + \frac{c}{\sqrt{n}}
\end{align}
for some $s \in (0,1)$, where $c$ is a finite non-negative constant.
\end{corollary}
\begin{proof}
    The inner bound follows directly by rewriting the error probability upper bounds in \eqref{eq:error_UB_0} and \eqref{eq:error_UB_1}. 
    The outer bound, on the other hand, can be shown to hold from the error probability lower bounds in \eqref{eq:error_LB_0} and \eqref{eq:error_LB_1}, as we demonstrate next.
    We start by restating a slightly loosened version of these bounds.

    Fix a channel discrimination decision region, associated with the tuple $(E_{0,n},E_{0,n})$. Then for every  $s \in (0,1)$, at least one of the following inequalities must hold
    \begin{align}
    \label{eq:En_UB_0}
       E_{0,n} & < D(P_s \| P_0 | \hat{P}_{x^n}) + \sqrt{\frac{2}{n} 
       \sum_{x\in\mathcal{X}} \hat{P}_{x^n}(x)\mu^{\prime \prime}_{x}(s)} + \frac{1}{n}\log 4  \\
       \label{eq:En_UB_1}
       E_{1,n} & < D(P_s \| P_1 | \hat{P}_{x^n}) + \sqrt{\frac{2}{n} 
       \sum_{x\in\mathcal{X}} \hat{P}_{x^n}(x)\mu^{\prime \prime}_{x}(s)} + \frac{1}{n}\log 4. 
    \end{align}
    Recalling that $\mu^{\prime \prime}_{x}(s)$ is the variance of the log-likelihood under the tilted distribution, we have
    \begin{align}
    \sum_{x\in\mathcal{X}} \hat{P}_{x^n}(x)\mu^{\prime \prime}_{x}(s)   \leq  \sum_{(x,z) \in \mathcal{X} \times \mathcal{Z}} \hat{P}_{x^n}(x) P_s(z|x) \left(\log \frac{P_1(z|x)}{P_0(z|x)}  \right)^2 \leq \max_{(x,z) \in \mathcal{X} \times \mathcal{Z}}  \left( \log \frac{P_1(z|x)}{P_0(z|x)}  \right)^2 
    \end{align}
     where the right-most upper bound is finite due to the assumption that $P_0(z|x)P_1(z|x) \neq 0$  for every  $z \in \mathcal{Z}$ and  $x \in \mathcal{X}$.
     Therefore, the bounds in \eqref{eq:En_UB_0} and \eqref{eq:En_UB_1} imply
    \begin{align}
       E_{0,n} & < D(P_s \| P_0 | \hat{P}_{x^n}) + \frac{c}{\sqrt{n}} \\
       E_{1,n} & < D(P_s \| P_1 | \hat{P}_{x^n}) +  \frac{c}{\sqrt{n}}
    \end{align}
    for some finite non-negative constant $c$.
    Now since at least one of these bounds must hold for every $s$, it follows that $0 \leq E_{0,n} \leq D(P_1 \| P_0 | \hat{P}_{x^n}) + \frac{c}{\sqrt{n}}$ and 
    $0 \leq E_{1,n} \leq D(P_0 \| P_1 | \hat{P}_{x^n}) + \frac{c}{\sqrt{n}}$.
    In the proof of Lemma \ref{lemma:Chernoff_identity} in Section \ref{section:minimax_and_NP}, we shall see that $D(P_s \| P_0 | \hat{P}_{x^n})$ is continuous and strictly increasing in $s \in [0,1]$, from $0$ to $D(P_1 \| P_0 | \hat{P}_{x^n})$.
    Therefore, there exists $s' \in [0,1]$ such that $E_{0,n}  = D(P_{s'} \| P_0 | \hat{P}_{x^n}) + \frac{c}{\sqrt{n}} $. For such a choice of $s'$, we must have $E_{1,n} \leq D(P_{s'} \| P_1 | \hat{P}_{x^n}) +  \frac{c}{\sqrt{n}}$. This concludes the proof.
\end{proof}
\section{Proof of Theorem \ref{theorem:rate_exponent_region}}
\label{sec:comm_disc_trade_off}
Equipped with the bounds on the error probabilities of channel discrimination, we now proceed to prove the result in Theorem \ref{theorem:rate_exponent_region}.
We start with the achievability and then present the converse. 
\subsection{Achievability}
We start with a channel coding achievability bound under a general input sequence set constraint. 
\begin{lemma}
\label{lemma:channel_coding_achievability}
Fix an input distribution $P_X \in \mathcal{P}(\mathcal{X})$ and block length $n$. Let $\mathcal{B}_n$ be a subset of $\mathcal{X}^n$. Then there exists a codebook $\mathcal{C}_n \subseteq \mathcal{B}_n$ of size $M_n$ such that for every $\tau > 0$, the average error probability satisfies
    \begin{equation}
        p_{\mathrm{e},n}^{\mathrm{av}} \times \Prb[X^n \in \mathcal{B}_n ]  \leq \Prb \left[ \imath(X^n ; Y^n) \leq n\tau + \log M_n \right] + \exp\{- n\tau\}
    \end{equation}
where $(X^n,Y^n) \sim P_X^n \times P_{Y|X}^n$, and  $\imath(x^n;y^n) = \log \frac{P^n_{Y|X}(y^n|x^n)}{P^n_Y(y^n)}$ is the information density.
\end{lemma}
The unconstrained version of the above result, i.e. with $\mathcal{B}_n  = \mathcal{X}^n $, was first derived by Shannon in \cite{Shannon1957} using a random coding argument.
In Appendix \ref{appendix:proof_lemma_rate}, we present a proof for the constrained version.
A similar yet stronger result, known as Feinstein's lemma, holds but for the maximal error probability, and is proved using Feinstein's maximal coding technique (see \cite{Polyanskiy_Wu_2024} for a modern treatment of these results).

We are now in a position to show that every tuple that lies in the region described in Theorem \ref{theorem:rate_exponent_region} is achievable in the sense of Definition \ref{definition:R_E_region}.
Let $(R,E_0,E_1)$ be one such tuple. This means that
\begin{align}
\label{eq:achievability_PX_s_conditions}
I(P_X,P_{Y|X}) \geq R, \ D\left( P_s \| P_0 | P_X \right) \geq E_0, \ D\left( P_s \| P_1 | P_X \right) \geq E_1, \ \text{and} \ E_{X \sim P_X} \left[b(X) \right]  \leq B
\end{align}
hold for some input distribution $P_X \in \mathcal{P}(\mathcal{X})$ and constant $s \in [0,1]$.
In what follows, we fix a block length $n$ and a pair $(P_X,s)$ sastifying \eqref{eq:achievability_PX_s_conditions}. We define the following subsets of input sequences
\begin{align}
\label{eq:sequence_constraint_set_0}
\mathcal{B}_{0,n}(P_X,s) & = \Big\{ x^n \in \mathcal{X}^n : 
 D(P_s \| P_0 | \hat{P}_{x^n} ) \geq D(P_s \| P_0| P_X) - \epsilon \Big\} \\
\label{eq:sequence_constraint_set_1}
\mathcal{B}_{1,n}(P_X,s) & = \Big\{ x^n \in \mathcal{X}^n : 
 D(P_s \| P_0 | \hat{P}_{x^n} ) \geq D(P_s \| P_0| P_X) - \epsilon \Big\} \\
\label{eq:sequence_constraint_set_b}
\mathcal{B}_{b,n}(P_X) & = \Big\{  x^n \in \mathcal{X}^n : \E_{X \sim \hat{P}_{x^n} }\left[b(X) \right] \leq \E_{X \sim P_X}[b(X)] + \epsilon    \Big\}.
\end{align}
for some small constant $\epsilon > 0$.
The inner bound in Corollary \ref{corollary:E_trade_off}, combined with \eqref{eq:achievability_PX_s_conditions}, \eqref{eq:sequence_constraint_set_0} and \eqref{eq:sequence_constraint_set_1}, imply that for every $x^n \in \mathcal{B}_{0,n}(P_X,s) \cap \mathcal{B}_{1,n}(P_X,s)$, there exists a decision region that satisfies\footnote{When more convenient, we highlight the dependency of $\varepsilon_{0,n}(x^n)$ and $\varepsilon_{1,n}(x^n)$ on $x^n$ instead of the message $w$.}
\begin{align}
- \frac{1}{n} \log \varepsilon_{0,n}(x^n) & \geq D(P_s \| P_0 | \hat{P}_{x^n} ) \geq D(P_s \| P_0| P_X) - \epsilon \geq E_0 - \epsilon  \\
 - \frac{1}{n} \log \varepsilon_{1,n}(x^n) & \geq D(P_s \| P_1 | \hat{P}_{x^n} ) \geq D(P_s \| P_1| P_X) - \epsilon \geq E_1 - \epsilon.
\end{align}
Therefore, by imposing the codebook constraint $\mathcal{C}_n \subseteq \mathcal{B}_n $, where
\begin{equation}
\mathcal{B}_n =  \mathcal{B}_{0,n}(P_X,s) \cap \mathcal{B}_{1,n}(P_X,s) \cap \mathcal{B}_{b,n}(P_X),
\end{equation}
the desired channel discrimination error exponents $(E_0,E_1)$ are achieved, while satisfying the input cost constraint.\footnote{Strictly speaking, and additional cost of $\epsilon$ is incurred. However, $\epsilon$ can be made to vanish as $n$ grows large.} It remains to show that the rate $R$ is also achievable. To do this, we use Lemma \ref{lemma:channel_coding_achievability} while setting $\frac{1}{n} \log M_n = I(P_X , P_{Y|X}) - (\tau + \delta)$ for some $\delta > 0$, from which we obtain 
\begin{equation}
p_{\mathrm{e},n}^{\mathrm{av}}  \leq \frac{\Prb \left[ \frac{1}{n}\imath(X^n ; Y^n) \leq I(P_X , P_{Y|X}) - \delta \right] + \exp \left\{ - n \tau \right\} }{\Prb[X^n \in \mathcal{B}_n ]}.
\end{equation}
Note that since $X^n \sim P_X^n$, then by the weak law of large numbers (WLLN) we have 
\begin{equation}
    \Prb \left[ \frac{1}{n}\imath(X^n ; Y^n) \leq I(P_X , P_{Y|X}) - \delta \right] \to 0, \ \text{as} \ n \to \infty.
\end{equation}
Similarly, by the WLLN, we also have 
\begin{equation}
\Prb \left[X^n \in \mathcal{B}_{0,n}(P_X,s) \right] \to 1, 
\ \Prb \left[X^n \in \mathcal{B}_{1,n}(P_X,s) \right] \to 1,
\ \Prb \left[X^n \in \mathcal{B}_{b,n}(P_X,s) \right] \to 1, \ \text{as} \ n \to \infty.
\end{equation}
Combining this with the fact that $\Prb \left[X^n \in \mathcal{B}_{n}\right]$ is lower bounded by
\begin{equation}
\Prb \left[X^n \in \mathcal{B}_{n}\right] \geq 1 - \Prb \left[X^n \notin \mathcal{B}_{0,n}(P_X,s) \right] - \Prb \left[X^n \notin \mathcal{B}_{0,n}(P_X,s) \right] - \Prb \left[X^n \notin \mathcal{B}_{0,n}(P_X,s) \right],
\end{equation}
it directly follows that $\Prb \left[X^n \in \mathcal{B}_{n}\right]  \to 1$ as $n \to \infty$, and therefore $p_{\mathrm{e},n}^{\mathrm{av}} \to 0$ as $n \to \infty$.
We conclude that for any $\epsilon > 0$, and by making $n$ large enough, there exists is a codebook of $M_n$ codewords in $\mathcal{B}_n$ such that
\begin{equation}
p_{\mathrm{e},n}^{\mathrm{av}} \leq \frac{1}{4}\epsilon 
\quad \text{and} \quad
\frac{1}{n}\log M_n = I(P_X,P_{Y|X}) - (\tau+\delta) \geq R - (\epsilon - \frac{1}{n} \log2).
\end{equation}
Note that the last inequality holds from \eqref{eq:achievability_PX_s_conditions}, and by choosing $\tau$, $\delta$ and $n$ appropriately.

Finally, using an expurgation argument, where we get rid of the worst half of the codewords (see, e.g, \cite[Ch. 7.7]{coverElementsInformationTheory2006}), it follows that there exists as codebook of size $M_n' = M_n/2$ in $\mathcal{B}_n$ for which $ \frac{1}{n} \log M_n' \geq R - \epsilon$ and  $p_{\mathrm{e},n} \leq 4 p_{\mathrm{e},n}^{\mathrm{av}} \leq \epsilon$. This concludes the proof of achievability.
\subsection{Converse}
\label{subsec:converse}
We now turn to the converse. First, for every $E_0$, $E_1$ and $B$, we define the following sets of distributions on $\mathcal{X}$ (i.e. subsets of $\mathcal{P}(\mathcal{X})$), which will prove useful further on
\begin{align}
\label{eq:set_P_E_s}
\mathcal{P}_E(\mathcal{X},E_0,E_1)  & = \bigcup_{s \in [0,1]} \left\{ P_X \in \mathcal{P}(\mathcal{X}) :  D\left( P_s \| P_0 | P_X \right) \geq E_0, \ D\left( P_s \| P_1 | P_X \right) \geq E_1 \right\} \\
\mathcal{P}_b(\mathcal{X},B)  & = \left\{ P_X \in \mathcal{P}(\mathcal{X}) :  
\E_{X \sim P_X} \left[ b(X)\right] \leq B \right\}.
\end{align}
Now suppose that the rate-exponent tuple $(R,E_0,E_1)$ is achievable. Then for every $\delta > 0$, and by making  $n$ large enough, there exists an $(n,M_n)$-code in which every codeword $x^n \in \mathcal{C}_n$ satisfies 
\begin{align}
\label{eq:E_0_converse}
E_0 - \delta & \leq -\frac{1}{n} \log \varepsilon_{0,n}(x^n) \leq  D(P_s \| P_0 \| \hat{P}_{x^n}) + \delta \\
\label{eq:E_1_converse}
E_1 - \delta & \leq  -\frac{1}{n} \log \varepsilon_{1,n}(x^n) \leq D(P_s \| P_1 \| \hat{P}_{x^n}) + \delta
\end{align}
for some parameter $s \in (0,1)$, which may depend on the codeword. The left-hand-side inequalities follow from Definition \ref{definition:R_E_region}, while the right-hand-side inequalities follow from the outer bound in Corollary \ref{corollary:E_trade_off}.
The constraints in \eqref{eq:E_0_converse} and \eqref{eq:E_1_converse}
imply that every codeword $x^n$ in $\mathcal{C}_n$ must have a type $\hat{P}_{x^n}$ that satisfies 
\begin{equation}
\label{eq:P_X_constraint}
    \hat{P}_{x^n} \in \mathcal{P}_E(\mathcal{X},E_0-\epsilon,E_1 - \epsilon) \cap \mathcal{P}_b(\mathcal{X},B)
\end{equation}
where $\epsilon = 2\delta$. Under this codeword constraint, our next goal is find an upper bound on the codebook size $M_n$, given the additional constraint $p_{\mathrm{e},n}(\mathcal{C}_n) \leq \epsilon$.
A challenge here is that the set of input distributions in \eqref{eq:P_X_constraint} is non-convex in general due to the union in \eqref{eq:set_P_E_s}. This prohibits us from applying existing approaches for proving channel coding converses under additive cost \cite{csiszarInformationTheoryCoding2011,ElGamal2011}, or additive sensing distortion \cite{Kobayashi2018,Ahmadipour2024,Zhang2011} constraints; as they crucially rely on the convexity of the underlying input distribution constraint set.

To circumvent this challenge, we first observe that $\mathcal{C}_n$ can be partitioned into constant composition sub-codebooks, each comprising codewords of the same type.
Let $P_X \in \mathcal{P}_n(\mathcal{X})$ be the type associate with the largest of such sub-codebooks, denoted by $\mathcal{C}_n(P_X )$, and let $M_n(P_X )$ be the corresponding number of codewords in this sub-codebook. 
It is straightforward to see
\begin{align}
\label{eq:const_comp_M}
 M_n \leq  |\mathcal{P}_n(\mathcal{X})| M_n(P_X) \leq (n+1)^{|\mathcal{X}|} M_n(P_X )
\end{align}
where the right-most inequality is due to \eqref{eq:type_counting_bound}.
Therefore, bounding the rate $\frac{1}{n} \log M_n$ of $\mathcal{C}_n$ is asymptotically equivalent to bounding the rate  $\frac{1}{n} \log M_n(P_X)$ of $\mathcal{C}_n(P_X )$, since $\frac{|\mathcal{X}|}{n} \log (n+1)$ tends to zero as $n$ approaches infinity.
Going forward, we focus on bounding $\frac{1}{n} \log M_n(P_X)$, which is more suited to our purpose as it can be done without relying on the convexity of the underlying input distribution constraint set.

Note that $p_{\mathrm{e},n}(\mathcal{C}_n) \leq \epsilon$ implies $p_{\mathrm{e},n}^{\mathrm{av}} (\mathcal{C}_n) \leq \epsilon$, which in turn implies $p_{\mathrm{e},n}^{\mathrm{av}} (\mathcal{C}_n(P_X )) \leq \epsilon$, which holds because $\mathcal{C}_n(P_X ) \subseteq \mathcal{C}_n$.
Therefore, from Fano's inequality (see, e.g., \cite[Section 3.1.4]{ElGamal2011}), 
we obtain
\begin{align}
\label{eq:Fano}
\frac{1}{n}\log M_n(P_X) \leq \frac{1}{n}I(X^n ; Y^n ) + \frac{1}{n} + \frac{\epsilon}{n}\log M_n(P_X)
\end{align}
where $X^n$ in \eqref{eq:Fano} is uniformly distributed on the constant composition sub-codebook $\mathcal{C}^n(P_X)$, since the undelying message in uniform.
We proceed by bounding the per-channel-use mutual information as
\begin{align}
\frac{1}{n}I(X^n ; Y^n ) &  \leq \frac{1}{n}\sum_{i = 1}^n I(X_i ; Y_i )   \\
& = \frac{1}{n} \sum_{i = 1}^n I(P_{X_i} , P_{Y|X} )  \\
\label{eq:converse_MI_UB_Jensen}
& \leq I(Q_X , P_{Y|X} ) 
\end{align}
where $P_{X_i}$ is obtained by marginalizing $P_{X^n}$, while $Q_{X}$ in \eqref{eq:converse_MI_UB_Jensen}  is given by the following mixture
\begin{equation}
    Q_X(x) = \frac{1}{n} \sum_{i=1}^n P_{X_i}(x), \ \forall x \in \mathcal{X}.
\end{equation}
The inequality in \eqref{eq:converse_MI_UB_Jensen} holds due to the concavity of the mutual information in the input distribution for a fixed channel and by Jensen's inequality. 
The next step is to show that the mixture input distribution $Q_{X}$ coincides with the sub-codebook's type $P_X$.
This is seen from the following 
\begin{align}
    Q_X(x) & = \frac{1}{n}  \sum_{i=1}^n \frac{1}{M_n(P_X)} \sum_{w = 1}^{M_n(P_X)} 
    \mathbbm{1} \left[ x_i(w) = x  \right] \\
    & = \frac{1}{M_n(P_X)} \sum_{w = 1}^{M_n(P_X)}   \frac{1}{n} \sum_{i=1}^n  \mathbbm{1} \left[ x_i(w) = x  \right] \\
    & = \frac{1}{M_n(P_X)}  \sum_{w = 1}^{M_n(P_X)}  \hat{P}_{x^n(w)}(x) \\
    \label{eq:type_equiv_Q_P}
    & = P_X(x).
\end{align}
Putting together \eqref{eq:type_equiv_Q_P}, \eqref{eq:converse_MI_UB_Jensen}, \eqref{eq:Fano} and \eqref{eq:const_comp_M}, we obtain 
\begin{align}
    \frac{1}{n} \log M_n & \leq \frac{1}{n} \log M_n(P_X) + \frac{|\mathcal{X}|}{n} \log(n+1)  \\
    & \leq I(P_X,P_{Y|X}) + \frac{1}{n} \big( 1 + \epsilon \log M_n(P_X) + \log(n+1) \big).
\end{align}
Since the rate $R$ is achievable, then $R - \delta \leq \frac{1}{n} \log M_n \leq I(P_X,P_{Y|X}) + \delta$ holds for sufficiently large $n$
Recall that $P_X$ must be in the set \eqref{eq:P_X_constraint}, from which we conclude that $(R,E_0,E_1)$ must satisfy
\begin{align}
     R & \leq I(P_X,P_{Y|X}) + \epsilon \\
     E_0 & \leq D(P_s \| P_0 | P_X) + \epsilon \\
     E_1 & \leq D(P_s \| P_1 | P_X) + \epsilon
\end{align}
for some $s \in [0,1]$ and $P_X \in \mathcal{P}_b(\mathcal{X},B)$.
These inequalities hold for every $\epsilon > 0$, as $\epsilon$ can be made as small as desired in the above proof by making $n$ sufficiently large. Therefore, we conclude that $(R,E_0,E_1)$ must satisfy the above inequalities while setting $\epsilon =0$. This concludes the converse proof.
\section{Minimax and Neyman-Pearson}
\label{section:minimax_and_NP}
In this section, we consider minimax and Neyman-Pearson metrics for channel discrimination
and characterize the resulting rate-exponent trade-offs. 
As we will see, the emerging trade-offs are special cases of the general trade-off in Theorem \ref{theorem:rate_exponent_region}.
\subsection{Minimax discrimination criterion}
In some applications, the two types of discrimination error probabilities are treated equally, and one wishes to control (i.e. minimize) the maximum of the two, defined for a given code as:
\begin{equation}
\varepsilon_{n}  =  \max \left\{ \varepsilon_{0,n}, \varepsilon_{1,n} \right\}.
\end{equation}
This gives rise to the following asymptotic trade-off.
\begin{definition}
\label{definition:R_E_region_minimax}
A tuple $(R,E)$ is said to be achievable under the minimax channel discrimination criterion if for every $\epsilon > 0$, and provided that $n$ is large enough, there exists an $(n,M_n)$-code with
\begin{equation}
\nonumber
p_{\mathrm{e},n} \leq \epsilon, \quad \frac{1}{n} \log  M_n \geq R - \epsilon  \quad \text{and} \quad 
\frac{1}{n} \log \frac{1}{\varepsilon_{n}} \geq E - \epsilon.
\end{equation}
The rate-exponent region $\mathcal{R}_{\mathrm{mini}}$ is the closure of the set of all achievable pairs $(R,E)$.
\end{definition}
It can be verified that $(R,E) \in \mathcal{R}_{\mathrm{mini}} $ if and only if 
$(R,E,E) \in \mathcal{R} $.
From this observation and Theorem \ref{theorem:rate_exponent_region}, it is readily seen that $\mathcal{R}_{\mathrm{mini}}$ is given by the set of all non-negative tuples $(R,E)$ such that 
\begin{align}
R & \leq I(P_X, P_{Y|X}) \\
\label{eq:E_minimax_before_chernoff}
E & \leq \max_{s\in [0,1]} \min\left\{ D(P_s\|P_0| P_X), D(P_s\|P_1 | P_X) \right\}
\end{align}
for some input distribution \(P_X \in \mathcal{P}(\mathcal{X}) \) that satisfies \(\E_{X \sim P_X}[b(X)] \leq B\).
Before we proceed, we define 
\begin{equation}
    C(P_0,P_1|P_X) \triangleq  - \min_{s \in [0,1]} \sum_{x \in \mathcal{X}}P_X(x)\log \sum_{z \in \mathcal{Z}}P_0(z|x)^{1-s} P_1(z|x)^{s} = - \min_{s \in [0,1]} \sum_{x \in \mathcal{X}}P_X(x) \mu_{x}(s)
\end{equation}
The above quantity is  a generalized form of the Chernoff information (see, e.g., \cite[Ch. 11.9]{coverElementsInformationTheory2006}), suited for channel discrimination. 
In the following result, we establish the equivalence between $C(P_0,P_1|P_X)$ and the right-hand-side of \eqref{eq:E_minimax_before_chernoff}, which parallels a known equivalence in classical hypothesis testing.
\begin{lemma}
\label{lemma:Chernoff_identity}
The following identity holds
\begin{equation}
    \max_{s\in [0,1]} \min\left\{ D(P_s\| P_0| P_X), D(P_s\| P_1 | P_X) \right\} = C(P_0,P_1|P_X)
\end{equation}
\end{lemma}
\begin{proof}
Taking the expectations of \eqref{eq:div_identity_x_1} and \eqref{eq:div_identity_x_2} with respect to $X \sim P_X$, we get
\begin{align}
    D\left(P_s\| P_0 | P_X \right) &= \sum_{x \in \mathcal{X}} P_X(x) \left(s\mu_x^{\prime}(s) - \mu_x(s) \right) = s\mu_{P_X}^{\prime}(s) - \mu_{P_X}(s)\\
     D\left(P_s \| P_1 | P_X \right) &= \sum_{x \in \mathcal{X}} P_X(x)  \left( (s-1)\mu_x^{\prime}(s)-\mu_x(s) \right) = (s-1)\mu_{P_X}^{\prime}(s) - \mu_{P_X}(s)
\end{align}
from which it follows directly that 
\begin{align}
\label{eq:D_s_derivative_0}
    \frac{\mathrm{d}}{\mathrm{d}s}D\left(P_s\| P_0 | P_X \right) &= \sum_{x \in \mathcal{X}} P_X(x) s \mu_x^{\prime \prime}(s) = s \mu_{P_X}^{\prime \prime}(s)\\
\label{eq:D_s_derivative_1}    
     \frac{\mathrm{d}}{\mathrm{d}s}D\left(P_s \| P_1 | P_X \right) &= \sum_{x \in \mathcal{X}} P_X(x)  (s-1) \mu_x^{\prime \prime}(s) =  (s-1) \mu_{P_X}^{\prime \prime}(s).
\end{align}
Recall from Section \ref{sec:channel_discrimination} that $\mu_x^{\prime \prime}(s) > 0$, unless the two channels $P_0(\cdot|x)$ and $P_1(\cdot|x)$ are identical under $x$.
If such inputs exist and $P_X$ is supported on them only, then we will have $ D(P_s\| P_0| P_X) =  D(P_s\| P_0| P_X) = 0$ for every $s \in [0,1]$, and hence $C(P_0,P_1|P_X) = 0$, and there is nothing left to prove. Therefore, we assume that $P_X(x) > 0$ and $\mu_x^{\prime \prime}(s) > 0$ for at least one $x$. 
From \eqref{eq:D_s_derivative_0} and \eqref{eq:D_s_derivative_1}, we see that $ D(P_s\|P_0| P_X)$ is strictly increasing in $s$, and $ D(P_s\|P_1| P_X)$ is strictly decreasing in $s$, and hence
\begin{equation}
    \min\left\{ D(P_s\|P_0| P_X), D(P_s\|P_1 | P_X) \right\} = \begin{cases} 
    D(P_s\|P_0 | P_X), & \text{if} \  D(P_s\|P_0| P_X) \leq  D(P_s\|P_1 | P_X) \\
    D(P_s\|P_1 | P_X), & \text{if} \  D(P_s\|P_0| P_X) >  D(W_s\|P_1 | P_X).
    \end{cases}
\end{equation}
It follows that the maximum is achieved at $s = s^{\star}$ that satisfies $  D(P_{ s^{\star}}\|P_0| P_X)= D(P_{s^{\star}}\|P_1 | P_X)$, 
for which we also have $\mu_{P_X}^\prime(s^{\star}) =0$.
Due to $\mu_{P_X}^{\prime \prime}(s) >0$ and $\mu_{P_X}(0) = \mu_{P_X}(1) =0$, it follows that
\begin{equation}
    \max_{s\in [0,1]} \min\left\{ D(P_s\|P_0| P_X), D(P_s\|P_1 | P_X) \right\} = - \min_{s \in [0,1]} \mu_{P_X}(s)
\end{equation}
which completes the proof of the lemma.
\end{proof}
We are now ready to present the rate-exponent trade-off under the minimax error criterion. 
\begin{theorem}
\label{theorem:minimax_region}
\(\mathcal{R}_{\mathrm{mini}}\) is given by the set of all non-negative pairs \((R,E)\) such that 
\begin{align}
R & \leq I(P_X, P_{Y|X}) \\
E & \leq C(P_0,P_1|P_X)
\end{align}
for some input distribution \(P_X \in \mathcal{P}(\mathcal{X})\) that satisfies \(\E_{X \sim P_X}[b(X)] \leq B\).
\end{theorem}
The proof follows directly from the above discussion.
\begin{example}
For the setting in Example \ref{example:1}, we have 
\begin{equation}
    C(P_0,P_1|P_X) = - \min_{s \in [0,1]} \rho \log \left( (1-q)^{1-s}q^s + q^{1-s}(1-q)^{s} \right) \label{eq:s_0.5}.
\end{equation}
Recall that $\E_{X\sim P_X}[\mu_{X}(s)] = \rho\log \left( (1-q)^{1-s}q^s + q^s(1-q)^{1-s} \right)$ is strictly convex in $s$. 
Since here we have $\E_{X\sim P_X}[\mu_{X}(s)] = \E_{X\sim P_X}[\mu_{X}(1-s)]$, it follows that $s=0.5$ is the  minimizer in \eqref{eq:s_0.5}, and we get
\begin{equation}
    C(P_0,P_1|P_X) = -\rho \log (2\sqrt{(1-q)q}) = -\rho \log e^{-D(0.5\| q)} = \rho D(0.5\| q).
\end{equation}
Therefore, $\mathcal{R}_{\mathrm{mini}}$ in this case is described by
\begin{align}
R & \leq H(\rho * p) - H(p) \\
E & \leq \rho D(0.5\| q)
\end{align}
for some $\rho \leq B$. This recovers the result in \cite[Theorem 1]{Joudeh2022}. As seen in Example \ref{example:1}, there is generally a trade-off between $R$ and $E$ whenever $B > 0.5$.
On the other hand, for the setting in Example \ref{example:2}, we have 
\begin{equation}
    C(P_0,P_1|P_X) = -\min_{s \in [0,1]} \log \left( (1-p_0)^{1-s}(1-p_1)^{1-s} +  p_0^sp_1^s\right).
\end{equation}
Since there is no trade-off between $R$ and $E$ in this case, the corresponding $\mathcal{R}_{\mathrm{mini}}$ is a rectangle.
\end{example}
\subsection{Neyman-Pearson discrimination criterion}
Here we consider the case where the two types of discrimination errors are treated unequally.
We adopt the Neyman-Pearson criterion, where the focus is on minimizing one type of error while keeping the other under control.
Here we choose to minimize the type II error probability while requiring that the type I error probability does not exceed a desired threshold $\alpha  \in (0,1)$. 

For a codebook $\mathcal{C}_n$ and given that the codeword $x^n$ has been sent, teh decision region  $\mathcal{A}_n(x^n)$ is chosen according to the above criterion, and the resulting type II discrimination error is given by
\begin{equation}
\beta_{\alpha,n}(x^n) \triangleq  \min_{\mathcal{A}_n(x^n) : \varepsilon_{0,n}(x^n) \leq \alpha}  \varepsilon_{1,n}(x^n).
\end{equation}
As argued in Section \ref{subsec:codes}, since it is not known beforehand which codeword in $\mathcal{C}_n$ will be sent, we take the maximum over all codewords in $\mathcal{C}_n$ and obtain an error probability  of
\begin{equation}
\beta_{\alpha,n} \triangleq  \max_{x^n \in \mathcal{C}_n} \beta_{\alpha,n}(x^n).
\end{equation}
Under the Neyman-Pearson criterion,  the asymptotic trade-off is formalized as follows.
\begin{definition}
\label{definition:R_E_region_NP}
Under the Neyman-Pearson discrimination criterion,  $(R,E_{\alpha})$ is said to be achievable if for every $\epsilon > 0$, and provided that $n$ is large enough, there exists an $(n,M_n)$-code with
\begin{equation}
\nonumber
p_{\mathrm{e},n} \leq \epsilon, \quad \frac{1}{n} \log  M_n \geq R - \epsilon  \quad \text{and} \quad 
\frac{1}{n} \log \frac{1}{\beta_{\alpha,n}} \geq E_{\alpha} - \epsilon.
\end{equation}
The rate-exponent region $\mathcal{R}_{\alpha}$ is the closure of the set of all achievable pairs $(R,E_{\alpha})$.
\end{definition}
We now present the trade-off under the Neyman-Pearson channel discrimination criterion. 
\begin{theorem}
\label{theorem:NP_region}
\(\mathcal{R}_{\alpha}\) for any $\alpha \in (0,1)$ is given by the set of all non-negative pairs  \((R,E)\) such that
    \begin{align}
        R & \leq I(P_X, P_{Y|X}) \\
        E & \leq D(P_0\|P_1|P_X) 
    \end{align}
for some input distribution \(P_X \in \mathcal{P}(\mathcal{X})\) that satisfies \(\E_{X \sim P_X}[b(X)] \leq B\).
\end{theorem}
The above theorem follows from Theorem \ref{theorem:rate_exponent_region}.
In particular, we know that the pair $(R,E_1)$ must satisfy $R \leq I(P_X, P_{Y|X}) $ and 
$E_1 \leq D(P_0\|P_1|P_X)$ for some \(P_X \in \mathcal{P}(\mathcal{X}) \) that satisfies \(\E_{X \sim P_X}[b(X)] \leq B\).
Achievability follows by choosing an arbitrarily small $s > 0$ in Theorem \ref{theorem:rate_exponent_region}, such that $s \to 0$ as $n \to \infty$. 
\begin{example}
For the setting in Example \ref{example:1}, the rate-exponent region $\mathcal{R}_{\alpha}$ is given by
\begin{align}
 R & \leq H(\rho * p) - H(p) \\
 E & \leq \rho d(q\|1- q)
\end{align}
for some $\rho \leq B$.
For the setting in Example \ref{example:2},  we have $R \leq H(\min\{0.5,B\} * p) - H(p)$ and $E \leq d(p_0\|p_1)$.
\end{example}

\section{Concluding Remarks}
We considered a problem of joint message communication and channel discrimination in discrete memoryless systems with an additive input cost, and we have established the optimal trade-off between the rate of reliable communication and the two types of channel discrimination error exponents. 
The simple instances given in Examples \ref{example:1}
and \ref{example:2} provide insights into the fundamental trade-off arising in JCAS systems, where the same resources are used simultaneously for both tasks.  

The setting we proposed in Fig. \ref{fig:block_diagram} can be extended to enable adaptivity, by modifying the encoding function such that at every channel use $i$, the input $X_i$ is made to be a function of both the message $W$ and previous sensor observations $Z^{i-1}$. 
However, it turns out that the results we have reported here will remain unchanged. This is because adaptivity does not improve the channel discrimination error exponents in block transmissions \cite{Hayashi2009,polyanskiy2011_ITA}, and will not improve the channel coding rate since the data channel does not depend on the parameter $\theta$.
This picture is different if we go beyond binary channel discrimination or if the data channel is made to depend on the parameter, as explored in \cite{Chang2022,Chang2023}. 

Our setting can be further extended in several other directions. 
For instance, it is of interest to extend the results to channels with general (e.g. continuous) alphabets. 
Although we do not expect the results to change dramatically, alternative proof techniques may be required. 
Note that unlike our previous preliminary works \cite{Wu2022,Wu2022arXiv} and the works of Chang et al.  \cite{Chang2022,Chang2023}, the achievability part in the current paper is directly applicable to channels with general alphabets; so is the converse part for channel discrimination.
Nevertheless, our channel coding rate converse proof relies on constant composition sub-codes and a type counting argument which, at least in its current form, only holds for discrete finite alphabets. 
It is of interest to try to extend the converse argument to general alphabets. 

Another extension is to consider, in addition to the coding rate and discrimination exponents, the channel coding exponent (i.e. reliability function) and to study the trade-off between all three. Some progress along these lines has been reported in \cite{Chang2023}.
It may also be of interest for practical purposes to derive refined bounds and second-order asymptotics, which tend to be tighter in finite blocklength regimes. 
Progress along these lines for the i.i.d. state model has been recently reported in \cite{Nikbakht2024}.
Finally, one may also consider extending the setup to incorporate variable-length sequential transmissions, which can result in significant gains for channel discrimination as recently reported in \cite{Chang2023sequential}.

\appendix
\section{Proof of Lemma \ref{lemma:channel_coding_achievability}}
\label{appendix:proof_lemma_rate}
The proof relies on random coding and threshold decoding (also known as Feinstein's decoding rule).
Define the following random coding distribution on $\mathcal{X}^n$
\begin{equation}
    Q_{X^n}(x^n) = \frac{P_X^n(x^n)}{P_X^n(\mathcal{B}^n)} \ind [ x^n \in \mathcal{B}_n ], \ \text{for all} \ x^n \in \mathcal{X}^n.
\end{equation}
Let $\bm{C}= \left\{X^n(1),X^n(2),\ldots,X^n(M_n)\right\}$ be a random codebook ensemble, in which codewords are independently drawn at random with distribution $Q_{X^n}$.
Since $Q_{X^n}$ is supported on $ \mathcal{B}_n$, every codebook in the random codebook ensemble satisfies the constraint that  codewords all belong to $ \mathcal{B}_n$.

For a fixed codebook 
$\bm{C} = \mathcal{C}_n$ and threshold $\gamma_n > 0$, and given a channel observation $Y^n = y^n$, the threshold decoder selects the unique message index $\hat{w} \in \mathcal{M}_n$ such that 
\begin{equation}
    \imath(x^n(\hat{w});y^n) = \log \frac{P_{Y|X}^n\big(x^n(\hat{w}) | y^n\big)}{P_Y^n(y^n)} > \gamma_n
\end{equation}
where $P_Y^n(y^n) = \E_{X^n \sim P_X^n} \left[ P_{Y|X}^n(y^n|X^n) \right]$. Note that the information density $\imath(x^n(\hat{w});y^n)$ is evaluated using $P_X^n$ as the input distribution, and not $Q_{X^n}$. 
This is done so that the final bound is in terms of $P_{Y|X}^n \times P_X^n$, which will become more apparent later in the proof.

The threshold decoder makes an error if the information density of the transmitted codeword does not exceed $\gamma_n$, or if any other codeword does. 
For a given codebook $\mathcal{C}_n$, we have
\begin{equation}
\nonumber
 p_{\mathrm{e},n}^{\mathrm{av}}(\mathcal{C}_n)  \leq \! \frac{1}{M_n}  \! \sum_{w \in \mathcal{M}_n} \! \!  \! \Big( \Prb \left[ \imath(x^n(w) ; Y^n) \leq \gamma_n  | W = w\right] + \Prb \left[ \imath(x^n(\bar{w}) ; Y^n) > \gamma_n, \ \text{for some} \ \bar{w} \neq w | W = w\right] \Big).
\end{equation}
Taking the expectation with respect to the codebook ensemble, we get
\begin{equation}
\label{eq:ensamble_average_error_ub}
\E \left[ p_{\mathrm{e},n}^{\mathrm{av}}(\bm{C}) \right] \leq \Prb \left[ \imath(X^n(1) ; Y^n) \leq \gamma_n | W = 1\right] + \sum_{\bar{w} \neq 1 }\Prb \left[ \imath(X^n(\bar{w}) ; Y^n) > \gamma_n | W = 1 \right]
\end{equation}
obtained by noting that codewords in $\bm{C}$ are independent and identically distributed, hence it suffices to restrict to $W =1$, and applying the union bound to get the second term on the right-hand-side.

We now treat the two terms on the right-hand-side of \eqref{eq:ensamble_average_error_ub} separately. The first is bounded as 
\begin{align}
    \Prb \left[ \imath(X^n(1) ; Y^n) \leq \gamma_n | W =1 \right] & = \sum_{x^n  \in \mathcal{X}^n} \sum_{y^n  \in \mathcal{Y}^n} Q_{X^n}(x^n) P^n_{Y|X}(y^n|x^n) \ind \left[ \imath(x^n ; y^n) \leq \gamma_n \right ]\\
     & \leq \sum_{x^n \in \mathcal{X}^n} \sum_{y^n  \in \mathcal{Y}^n} \frac{P_X^n(x^n)}{P_X^n(\mathcal{B}^n)} P^n_{Y|X}(y^n|x^n) \ind \left[ \imath(x^n ; y^n) \leq \gamma_n \right] \\
     & = \frac{\Prb \left[ \imath(X^n ; Y^n) \leq \gamma_n \right]}{\Prb[X^n \in \mathcal{B}_n ]}
\end{align}
where in the last line, we have $(X^n,Y^n) \sim P_X^n \times P_{Y|X}^n$.
As for the second term, for any $\bar{w} \neq 1$, we have 
\begin{align}
\Prb \left[ \imath(X^n(\bar{w}) ; Y^n) > \gamma_n | W = 1 \right] & = \sum_{x^n \in \mathcal{X}^n}  \sum_{y^n \in \mathcal{Y}^n} Q_{X^n}(x^n)Q_{Y^n}(y^n) \ind \left[ \frac{P_{Y|X}^n(y^n|x^n)}{P_Y^n(y^n)} > \exp\{\gamma_n\} \right] \\
\label{eq:P_X_w_bar_UB}
& \leq \exp\{-\gamma_n\} \sum_{x^n \in \mathcal{X}^n} \sum_{y^n \in \mathcal{Y}^n} Q_{X^n}(x^n)Q_{Y^n}(y^n) \times \frac{P_{Y|X}^n(y^n|x^n)}{P_Y^n(y^n)}
\end{align}
where $Q_{Y^n}(y^n) = \E_{X^n \sim Q_{X^n}} \left[ P_{Y|X}^n(y^n|X^n) \right]$.
We bound the product $Q_{X^n}(x^n)Q_{Y^n}(y^n)$ as follows
\begin{align}
Q_{X^n}(x^n)Q_{Y^n}(y^n) & = Q_{X^n}(x^n) \E_{X^n \sim Q_{X^n}} \left[ P^n_{Y|X}(y^n|X^n) \right] \\
    & = \frac{\ind \left[ x^n \in \mathcal{B}_n \right] }{P_X^n(\mathcal{B}^n)} P_X^n(x^n) \E_{X^n \sim P_X^n} \left[ P^n_{Y|X}(y^n|X^n)  \frac{\ind \left[ X^n \in \mathcal{B}_n \right] }{P_X^n(\mathcal{B}_n )}\right]  \\
    & \leq \frac{\ind \left[ x^n \in \mathcal{B}_n \right] }{P_X^n(\mathcal{B}^n)} P_X^n(x^n) \E_{X^n \sim P_X^n} \left[ P^n_{Y|X}(y^n|X^n) \right] \\
    &  =\frac{\ind \left[x^n \in \mathcal{B}_n\right] }{\big(P_X^n(\mathcal{B}^n)\big)^2} P_X^n(x^n) P^n_{Y}(y^n).
\end{align}
Substituting this into \eqref{eq:P_X_w_bar_UB}, we obtain the following. 
\begin{align}
\Prb \left[ \imath(X^n(\bar{w}) ; Y^n) > \gamma_n | W = 1 \right]& \leq \frac{\exp\{-\gamma_n\}}{\big(P_X^n(\mathcal{B}^n)\big)^2} \sum_{x^n \in \mathcal{X}^n}  P_X^n(x^n) \ind \left[ x^n \in \mathcal{B}_n \right]  \sum_{y^n \in \mathcal{Y}^n}P_{Y|X}^n(y^n|x^n) \\
& = \frac{\exp\{-\gamma_n\}}{P_X^n(\mathcal{B}^n)}.
\end{align}
Putting everything together, we get the following upper bound
\begin{equation}
\E \left[ p_{\mathrm{e},n}^{\mathrm{av}}(\bm{C}) \right] \leq \frac{ \Prb \left[ \imath(X^n ; Y^n) \leq \gamma_n \right] + (M_n -1 )\exp\{-\gamma_n\} }{\Prb[X^n \in \mathcal{B}_n ]}.
\end{equation}
The final result is obtained by loosening $M_n - 1$ to $M_n$, setting $\gamma_n = n\tau + \log M_n$, and from the fact that there must exist at least one codebook in the ensemble for which the upper bound holds. 
\section*{Acknowledgements}
This work is funded in part by the European Union (ERC, IT-JCAS, 101116550). Views and opinions expressed are however those of the authors only and do not necessarily reflect those of the European Union or the European Research Council. Neither the European Union nor the granting authority can be held responsible for them. Han Wu would like to thank Prof. Giuseppe Durisi for introducing hypothesis testing to him when he was studying at Chalmers University of Technology.
\bibliographystyle{IEEEtran}
\bibliography{References}

\begin{thebibliography}{10}
\providecommand{\url}[1]{#1}
\csname url@samestyle\endcsname
\providecommand{\newblock}{\relax}
\providecommand{\bibinfo}[2]{#2}
\providecommand{\BIBentrySTDinterwordspacing}{\spaceskip=0pt\relax}
\providecommand{\BIBentryALTinterwordstretchfactor}{4}
\providecommand{\BIBentryALTinterwordspacing}{\spaceskip=\fontdimen2\font plus
\BIBentryALTinterwordstretchfactor\fontdimen3\font minus \fontdimen4\font\relax}
\providecommand{\BIBforeignlanguage}[2]{{%
\expandafter\ifx\csname l@#1\endcsname\relax
\typeout{** WARNING: IEEEtran.bst: No hyphenation pattern has been}%
\typeout{** loaded for the language `#1'. Using the pattern for}%
\typeout{** the default language instead.}%
\else
\language=\csname l@#1\endcsname
\fi
#2}}
\providecommand{\BIBdecl}{\relax}
\BIBdecl

\bibitem{Liu2022}
F.~Liu, Y.~Cui, C.~Masouros, J.~Xu, T.~X. Han, Y.~C. Eldar, and S.~Buzzi, ``Integrated sensing and communications: Toward dual-functional wireless networks for {6G} and beyond,'' \emph{IEEE Journal on Selected Areas in Communications}, vol.~40, no.~6, pp. 1728--1767, 2022.

\bibitem{Xiong2024}
Y.~Xiong, F.~Liu, K.~Wan, W.~Yuan, Y.~Cui, and G.~Caire, ``From torch to projector: Fundamental tradeoff of integrated sensing and communications,'' \emph{IEEE BITS the Information Theory Magazine}, pp. 1--13, 2024.

\bibitem{Wild2021}
T.~Wild, V.~Braun, and H.~Viswanathan, ``Joint design of communication and sensing for beyond {5G} and {6G} systems,'' \emph{IEEE Access}, vol.~9, pp. 30\,845--30\,857, 2021.

\bibitem{Sturm2011}
C.~Sturm and W.~Wiesbeck, ``Waveform design and signal processing aspects for fusion of wireless communications and radar sensing,'' \emph{Proc. IEEE}, vol.~99, no.~7, pp. 1236--1259, 2011.

\bibitem{Ma2020}
D.~Ma, N.~Shlezinger, T.~Huang, Y.~Liu, and Y.~C. Eldar, ``Joint radar-communication strategies for autonomous vehicles: Combining two key automotive technologies,'' \emph{IEEE Signal Process. Magazine}, vol.~37, no.~4, pp. 85--97, 2020.

\bibitem{Kobayashi2018}
M.~Kobayashi, G.~Caire, and G.~Kramer, ``Joint state sensing and communication: Optimal tradeoff for a memoryless case,'' in \emph{IEEE International Symposium on Information Theory (ISIT)}, 2018, pp. 111--115.

\bibitem{Sutivong2005}
A.~Sutivong, M.~Chiang, T.~Cover, and Y.-H. Kim, ``Channel capacity and state estimation for state-dependent gaussian channels,'' \emph{IEEE Transactions on Information Theory}, vol.~51, no.~4, pp. 1486--1495, 2005.

\bibitem{Choudhuri2013}
C.~Choudhuri, Y.-H. Kim, and U.~Mitra, ``Causal state communication,'' \emph{IEEE Transactions on Information Theory}, vol.~59, no.~6, pp. 3709--3719, 2013.

\bibitem{Zhang2011}
W.~Zhang, S.~Vedantam, and U.~Mitra, ``Joint transmission and state estimation: A constrained channel coding approach,'' \emph{IEEE Transactions on Information Theory}, vol.~57, no.~10, pp. 7084--7095, 2011.

\bibitem{Kobayashi2019}
M.~Kobayashi, H.~Hamad, G.~Kramer, and G.~Caire, ``Joint state sensing and communication over memoryless multiple access channels,'' in \emph{IEEE International Symposium on Information Theory (ISIT)}, 2019, pp. 270--274.

\bibitem{Ahmadipour2024}
M.~Ahmadipour, M.~Kobayashi, M.~Wigger, and G.~Caire, ``An information-theoretic approach to joint sensing and communication,'' \emph{IEEE Transactions on Information Theory}, vol.~70, no.~2, pp. 1124--1146, 2024.

\bibitem{Ahmadipour2023}
M.~Ahmadipour and M.~Wigger, ``An information-theoretic approach to collaborative integrated sensing and communication for two-transmitter systems,'' \emph{IEEE Journal on Selected Areas in Information Theory}, vol.~4, pp. 112--127, 2023.

\bibitem{Ahmadipour2023b}
M.~Ahmadipour, M.~Wigger, and S.~Shamai, ``Integrated communication and receiver sensing with security constraints on message and state,'' in \emph{IEEE International Symposium on Information Theory (ISIT)}, 2023, pp. 2738--2743.

\bibitem{Gunlu2023}
O.~Günlü, M.~R. Bloch, R.~F. Schaefer, and A.~Yener, ``Secure integrated sensing and communication,'' \emph{IEEE Journal on Selected Areas in Information Theory}, vol.~4, pp. 40--53, 2023.

\bibitem{Xiong2023}
Y.~Xiong, F.~Liu, Y.~Cui, W.~Yuan, T.~X. Han, and G.~Caire, ``On the fundamental tradeoff of integrated sensing and communications under {Gaussian} channels,'' \emph{IEEE Transactions on Information Theory}, vol.~69, no.~9, pp. 5723--5751, 2023.

\bibitem{Joudeh2024}
H.~Joudeh and G.~Caire, ``Joint communication and state sensing under logarithmic loss,'' in \emph{4th IEEE International Symposium on Joint Communications \& Sensing (JC\&S)}, 2024, pp. 1--6.

\bibitem{Joudeh2022}
H.~Joudeh and F.~M.~J. Willems, ``Joint communication and binary state detection,'' \emph{IEEE Journal on Selected Areas in Information Theory}, vol.~3, no.~1, pp. 113--124, 2022.

\bibitem{Wu2022}
H.~Wu and H.~Joudeh, ``On joint communication and channel discrimination,'' in \emph{IEEE International Symposium on Information Theory (ISIT)}, 2022, pp. 3321--3326.

\bibitem{Wu2022arXiv}
------, ``Joint communication and channel discrimination,'' \emph{arXiv:2208.07450v1}, 2022.

\bibitem{Chang2022}
M.-C. Chang, T.~Erdogan, S.-Y. Wang, and M.~R. Bloch, ``Rate and detection error-exponent tradeoffs of joint communication and sensing,'' in \emph{2nd IEEE International Symposium on Joint Communications \& Sensing (JC\&S)}, 2022, pp. 1--6.

\bibitem{Chang2023}
M.-C. Chang, S.-Y. Wang, T.~Erdoğan, and M.~R. Bloch, ``Rate and detection-error exponent tradeoff for joint communication and sensing of fixed channel states,'' \emph{IEEE Journal on Selected Areas in Information Theory}, vol.~4, pp. 245--259, 2023.

\bibitem{Joudeh2023}
H.~Joudeh, ``Joint communication and target detection with multiple antennas,'' in \emph{26th International ITG Workshop on Smart Antennas and 13th Conference on Systems, Communications, and Coding}, 2023, pp. 1--6.

\bibitem{Ahmadipour2023strong}
M.~Ahmadipour, M.~Wigger, and S.~Shamai, ``Strong converses for memoryless bi-static {ISAC},'' in \emph{IEEE International Symposium on Information Theory (ISIT)}, 2023, pp. 1818--1823.

\bibitem{Ahmadipour2024strong}
------, ``Strong converse for bi-static {ISAC} with two detection-error exponents,'' in \emph{International Zurich Seminar on Information and Communication (IZS)}, 2024, p.~45.

\bibitem{Weinberger2014}
N.~Weinberger and N.~Merhav, ``Codeword or noise? exact random coding exponents for joint detection and decoding,'' \emph{IEEE Transactions on Information Theory}, vol.~60, no.~9, pp. 5077--5094, 2014.

\bibitem{Weinberger2017}
------, ``Channel detection in coded communication,'' \emph{IEEE Transactions on Information Theory}, vol.~63, no.~10, pp. 6364--6392, 2017.

\bibitem{chernoff1952}
H.~Chernoff, ``A measure of asymptotic efficiency for tests of a hypothesis based on the sum of observations,'' \emph{Ann. Math. Stat.}, pp. 493--507, 1952.

\bibitem{hoeffding1965}
W.~Hoeffding, ``Asymptotically optimal tests for multinomial distributions,'' \emph{Ann. Math. Stat.}, pp. 369--401, 1965.

\bibitem{shannonLowerBoundsError1967}
C.~E. Shannon, R.~G. Gallager, and E.~R. Berlekamp, ``Lower bounds to error probability for coding on discrete memoryless channels. {{I}},'' \emph{Information and Control}, vol.~10, no.~1, pp. 65--103, Jan. 1967.

\bibitem{csiszar1971}
I.~Csisz{\'a}r and G.~Longo, ``On the error exponent for source coding and for testing simple statistical hypotheses,'' \emph{Studia Sci. Math. Hungar.}, vol.~6, pp. 181--191, 1971.

\bibitem{blahutHypothesisTestingInformation1974}
R.~Blahut, ``Hypothesis testing and information theory,'' \emph{IEEE Transactions on Information Theory}, vol.~20, no.~4, pp. 405--417, Jul. 1974.

\bibitem{Hayashi2009}
M.~Hayashi, ``Discrimination of two channels by adaptive methods and its application to quantum system,'' \emph{IEEE Transactions on Information Theory}, vol.~55, no.~8, pp. 3807--3820, 2009.

\bibitem{polyanskiy2011_ITA}
Y.~Polyanskiy and S.~Verd{\'u}, ``Binary hypothesis testing with feedback,'' in \emph{Inf. Theory Appl. Workshop (ITA)}, 2011.

\bibitem{Nitinawarat2013}
S.~Nitinawarat, G.~K. Atia, and V.~V. Veeravalli, ``Controlled sensing for multihypothesis testing,'' \emph{IEEE Transactions on Automatic Control}, vol.~58, no.~10, pp. 2451--2464, 2013.

\bibitem{csiszarInformationTheoryCoding2011}
I.~Csisz{\'a}r and J.~K{\"o}rner, \emph{Information {{Theory}}: Coding {{Theorems}} for {{Discrete Memoryless Systems}}}, 2nd~ed.\hskip 1em plus 0.5em minus 0.4em\relax {Cambridge}: {Cambridge University Press}, 2011.

\bibitem{ElGamal2011}
A.~El~Gamal and Y.-H. Kim, \emph{Network Information Theory}.\hskip 1em plus 0.5em minus 0.4em\relax {Cambridge ; New York}: {Cambridge University Press}, 2011.

\bibitem{Shannon1957}
C.~E. Shannon, ``Certain results in coding theory for noisy channels,'' \emph{Information and Control}, vol.~1, no.~1, pp. 6--25, 1957.

\bibitem{Polyanskiy_Wu_2024}
Y.~Polyanskiy and Y.~Wu, \emph{Information Theory: From Coding to Learning}.\hskip 1em plus 0.5em minus 0.4em\relax Cambridge University Press, 2024.

\bibitem{coverElementsInformationTheory2006}
T.~M. Cover and J.~A. Thomas, \emph{Elements of Information Theory}, 2nd~ed.\hskip 1em plus 0.5em minus 0.4em\relax {Hoboken, N.J}: {Wiley-Interscience}, 2006.

\bibitem{Nikbakht2024}
H.~Nikbakht, M.~Wigger, S.~Shamai, and H.~V. Poor, ``Integrated sensing and communication in the finite blocklength regime,'' \emph{arXiv:2401.15752}, 2024.

\bibitem{Chang2023sequential}
M.-C. Chang, S.-Y. Wang, and M.~R. Bloch, ``Sequential joint communication and sensing of fixed channel states,'' in \emph{IEEE Information Theory Workshop (ITW)}.\hskip 1em plus 0.5em minus 0.4em\relax IEEE, 2023, pp. 462--467.

\end{thebibliography}
\end{document}